\documentclass[useAMS,usenatbib,usegraphicx]{mn2e}

\title[The molecular polar disc in NGC~2768]{The molecular polar disc in NGC~2768\thanks{Based on observations carried out with the IRAM Plateau de Bure
Interferometer. IRAM is supported by INSU/CNRS (France), MPG (Germany)
and IGN (Spain).}}
\author[A. F. Crocker et al.]{Alison F. Crocker$^{1}$, Martin
  Bureau$^{1}$, Lisa M. Young$^{1,2}$, Francoise Combes$^{3}$\\
$^{1}$Sub-Department of Astrophysics, University of Oxford, Denys
  Wilkinson Building, Keble Road, Oxford OX1 3RH \\
$^{2}$Physics Department, New Mexico Institute of Mining and
Technology, Socorro, NM 87801, U.S.A. \\
$^{3}$Observatoire de Paris, LERMA, 61 Av. de l'Observatoire, 75014,
Paris, France}

\begin{document}

\date{}

\pagerange{\pageref{firstpage}--\pageref{lastpage}} \pubyear{}

\maketitle

\label{firstpage}

\begin{abstract}

We present CO(1-0) and CO(2-1) maps of the molecular polar disc in
the elliptical
galaxy NGC~2768 obtained at the IRAM Plateau de Bure
Interferometer. The maps have a resolution of $2\farcs6 \times
2\farcs3$ and $1\farcs2 \times 1\farcs2$ for the CO(1-0) and CO(2-1)
lines, respectively. The CO maps complete the unique picture of the
interstellar medium (ISM) of NGC~2768; the dust, molecular gas, ionised gas
and neutral hydrogen (H$\:${\small I}) trace 
the recent acquisition of cold and cool gas over two orders of
magnitude in radii (and much more in density). In agreement with the other ISM components, the CO distribution extends nearly
perpendicularly to the photometric major axis of the galaxy. Velocity maps of the CO
show a rotating polar disc or ring in the inner kiloparsec. This cool gas could lead to
kinematic substructure formation within NGC~2768. However, the stellar velocity
field and H$\beta$ absorption
linestrength maps from the optical integral-field spectrograph SAURON give no
indication of a young and dynamically cold stellar population
coincident with the molecular polar disc. Very recent
or weak 
star formation, undetectable in linestrengths, nevertheless remains a possibility
and could be at the origin of some of the ionised gas observed. Millimetre
continuum emission was also detected in NGC~2768, 
now one of only a few low-luminosity active galactic nuclei with observed millimetre
continuum emission.

\end{abstract}

\begin{keywords}
galaxies: individual: NGC2768 -- galaxies: elliptical and lenticular,
cD -- galaxies: ISM -- galaxies: evolution -- galaxies: kinematics and
dynamics -- galaxies: structure
\end{keywords}

\section{Introduction}

Galaxy formation scenarios must explain the rather uniform global
properties of early-type galaxies as seen in the fundamental plane
\citep[e.g.][]{dressler87,djorgovski87} 
while simultaneously accounting for the diversity seen in the internal
kinematics and stellar populations \citep[e.g.][]{dezeeuw02}. 
Indeed, the maps of the stellar kinematics of 48 elliptical and S0
galaxies from the SAURON optical integral-field unit (IFU) have shown a wide
variety of kinematic 
substructures, from large embedded discs to small counter-rotating cores 
\citep[e.g.][]{emsellem04,mcdermid06}. Major mergers have been found to
reproduce many of the kinematic substructures observed in the SAURON
maps without invoking post-merger star formation \citep{J}. However,
many of the kinematically decoupled components (KDCs) in the SAURON
maps are very young \citep{kuetal06,mcdermid06}, suggesting their
stars formed in situ, and not as part of another galaxy. 

Over the past twenty years, it has been shown that E/S0s do have
detectable amounts of cool and cold gas. Cold dust was first found in
the IRAS survey in which around 45\% of ellipticals were detected in
both the 60 and 100~$\mu$m bands \citep{knapp}. Molecular gas detections
have often been biased toward dust-rich or infrared-bright galaxies, but
two recent surveys do not select on these criteria and give detection
rates of 28\% for E/S0s in the SAURON representative sample
\citep*{CYB07} and 78\% for S0s in a volume-limited sample
\citep{sage06}. Particularly interesting for substructure formation are
the maps of cold molecular gas provided by interferometric CO
observations \citep[e.g.][]{young02, young05}. These maps often show rotating
molecular gas discs on the same spatial scales as KDCs. 

While the current evidence indirectly connects molecular gas discs
with young KDCs, the details of
the connection are still unknown. Do all molecular discs form stellar
components? Do stars form everywhere in the molecular disc, or only in
the densest regions? Does the regularity of the morphology of the
molecular disc (presumably indicating how long since the gas was
accreted) correlate with the age of the stars? To answer these
questions, we must compare the molecular gas morphology and kinematics
to the stellar morphology, kinematics and ages in a
spatially-resolved manner. This comparison requires
interferometric CO maps and optical (or near-infrared) IFU data for
many individual
galaxies. With this comparison in mind, we have started to
obtain CO synthesis maps for the $12$ early-type SAURON galaxies
recently detected in CO \citep[see][and references therein]{CYB07}. 

In this paper, we present the first CO synthesis maps of NGC~2768 and
compare the molecular gas to the stellar and ionised gas properties
from SAURON. NGC~2768 is classified as an E6 galaxy in
RC3 \citep{devaucouleurs91}, an S0 in the Carnegie Atlas of Galaxies
\citep{sandage94}, and as a fast
rotator in the recently
proposed kinematic classification scheme of \citet{em07}. \citet{kim89} first
discovered both the
polar orientation of the inner dust lane and the rotation of the ionized gas
about the major axis, suggesting an external accretion origin for
the dust and gas. NGC~2768 has a low-luminosity active galactic
nucleus (AGN) with a low-ionisation nuclear emission-line region (LINER) 
spectrum \citep{heckman80}, a compact radio core \citep*{nagar05} and
an X-ray
source consistent with being a point source \citep*{komossa99}. General
properties of NGC~2768 are listed in Table~\ref{tab:basic}. 

This paper is organized as follows. In Section 2, we describe the calibration and 
reduction of the data from the Plateau de Bure Interferometer. Section
3 presents 
flux measurements for the continuum data  and channel maps, total
intensity maps and velocity fields for the CO line data. In Section 4, we
discuss the origin and fate of the molecular gas and also consider the
contributions to the mm continuum flux from the low-luminosity AGN
(LLAGN) and cold dust. We draw our conclusions in Section 5.

\begin{table}
 \caption{Basic properties of NGC~2768. The left and middle columns
 list the different quantities and their values; the right column
 lists corresponding references.}
 \begin{tabular}{@{}llr} 
  \hline
  Quantity & Value & Ref. \\
  \hline
  R.A. (J2000.0) & 09 11 37.413 & 1\\
  Dec. (J2000.0) & +60 02 14.86 & 1\\
  Heliocentric Velocity & $1373\pm5$ km s$^{-1}$ & 2\\
  Distance & $21.5$ Mpc & 3\\
  Scale & 1\arcsec $=$ 104 pc & 3\\
  Type (RC3) & E6 & 4\\
  Type (CAG) & S0 & 5\\
  Corrected apparent B mag & $10.60$ & 6\\
  Corrected absolute B mag & $-21.15$ & 6 \\
  L$_{\mathrm{B}}$ & $2.3 \times 10^{10}$ L$_{\sun}$ & 7\\
  L$_{\mathrm{FIR}}$ & $4.3\times10^{8}$ L$_{\sun}$ & 7\\
  L$_{\mathrm{FIR}}$/L$_{\mathrm{B}}$ & $1.9\times10^{-2}$ & 7\\
  L$_{\mathrm{FIR}}$/M$_{\mathrm{H}_{2}}$ & $6.3$ L$_{\sun}$/M$_{\sun}$ & 7\\ 

  \hline
 \end{tabular}
\label{tab:basic}

References: (1) \citet{nagar05}; (2) NED; (3) \citet{emsellem04};
 (4) \citet{devaucouleurs91}; (5) \citet{sandage94}; (6) HyperLEDA;
(7) Derived quantity using data from NED, IRAS \citep{moshir90}, and this paper.
\end{table}

\section[]{Observations and Data Reduction}

We observed NGC~2768 in December 2005 (C configuration) and August 2006
(D configuration) at the IRAM Plateau de Bure Interferometer (PdBI). 
The $^{12}$CO(1-0) and $^{12}$CO(2-1) lines were observed
simultaneously using dual-band SIS receivers. All six 15m antennae
were used for the C-configuration observations in December while only
five antennae were available for the August observations. Typical
system temperatures were 200-300 K for the CO(1-0) line in both
months and the CO(2-1) in December. In the worse weather of August,
the CO(2-1) line observations had system temperatures of 300-1000 K. The spectral
correlators were centred at 114.743 GHz and 229.482 GHz, the 
transition frequencies of CO(1-0) and CO(2-1) corrected for the
galaxy's heliocentric velocity of 1373 km s$^{-1}$. The correlator
configuration used four 160 MHz-wide (417 and 209 km s$^{-1}$ for
CO(1-0) and CO(2-1), respectively) units with a frequency resolution of 1.25 MHz
(3.3 and 1.6 km s$^{-1}$), covering a total bandwidth of 580 MHz (1670
and 835 km s$^{-1}$). The correlator was regularly calibrated by a 
noise source inserted in the IF system.  

We obtained visibilities with series of twenty 1 min integrations on
source, followed by three $45$ s phase and amplitude calibrations on a 
nearby quasar, 0836+710. As 0836+710 was not bright enough, we used
3C84 (25 December), 0528+134 (25 August) and MWC349 (28 August) as flux
calibrators. Comparing the flux measured for 0836+710 on 28 August to
that measured on 25 December suggests an uncertainty in our flux
calibration of $\approx20$\%. This is unfortunately not a very good
estimate as the true flux of 0836+710 may have varied over the 8 month
interval.   

The data were reduced with the Grenoble Image and Line Data Analysis
System (GILDAS) software packages CLIC and MAPPING \citep{GL}.
Using CLIC, we first calibrated the data using
the standard pipeline. Data obtained during periods of bad weather were
flagged and then ignored.  After calibration, we used MAPPING to
create two data cubes for both lines with velocity planes separated by 40 km
s$^{-1}$ and 60 km s$^{-1}$. The primary beam size is $44\arcsec$ for the
CO(1-0) observations and $22\arcsec$ for the CO(2-1) observations. We
choose the spatial dimensions of the datacube to be twice the diameter of the
primary beam, $88\arcsec\times88\arcsec$ and
$44\arcsec\times44\arcsec$ for the CO(1-0) and C(2-1) lines,
respectively.  The synthesized beam sizes are $2\farcs6 \times
2\farcs3$ for CO(1-0) and $1\farcs2 \times 1\farcs2$ for CO(2-1). We thus choose spatial pixels of
$.7\arcsec\times.7\arcsec$ and $.35\arcsec\times.35\arcsec$ for the
CO(1-0) and CO(2-1) lines, respectively. As we detected continuum
emission, we subtracted the continuum using the task contsub in the
Multichannel Image Reconstruction, Image Analysis and Display (MIRIAD)
software package \citep{sault95}. The dirty beam has small
side lobes that necessitated cleaning the datacube. The cleaning was
done using the H\"ogbom method \citep{hogbom74}; we stopped cleaning
in each velocity plane after the brightest 
residual pixel had a value lower than the rms of the 
uncleaned datacube. 

To constrain the continuum emission, we selected channels at least 40
km s$^{-1}$ away from the highest and lowest velocity channels with any
line emission in the cleaned data cubes.
The very edges of the 580 MHz bandwidth were also
avoided. The resulting continuum windows were 356 and 200 MHz wide at 115
and 230 GHz, respectively. We used the MAPPING tasks
with the same parameters as for the line data to create
clean images of the continuum at both frequencies. We fit point-source models
in the $uv$ plane to find the flux of the 3 mm continuum and an upper limit for
the 1 mm continuum. The continuum emission is spatially unresolved at
both frequencies. 

\section[] {Results}

We detected extended emission in both the CO(1-0) and CO(2-1) lines and
point-source continuum emission at 3 mm. 

\subsection[] {Continuum emission}

\begin{figure}
\begin{center}
\rotatebox{0}{\includegraphics[width=8cm]
  {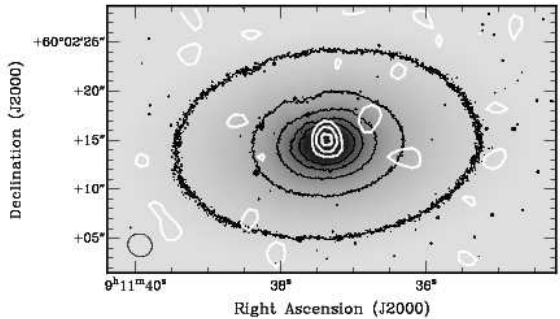}}\\
\vspace*{3mm}
\caption{3 mm continuum contours (white) over a Hubble Space Telescope (HST)
  Advanced Camera for Surveys (ACS) image (F814W
  filter, black contours) of NGC~2768. Contours are at 2, 4, and 6$\sigma$, where $\sigma=0.28$
  mJy beam$^{-1}$. The synthesized beam is plotted in the lower left corner.}
\label{fig:cont}
\end{center}
\end{figure}

The continuum maps at 1 and 3 mm both show peaks at the centre of
NGC~2768. The 3 mm continuum map has a peak of over $6\sigma$ consistent with
being a point source (see Fig.~\ref{fig:cont}). Fitting a
point-source model in the $uv$-plane yields a flux density of  1.94
mJy at a right ascension of 09 11 37.38 and declination of +60 02
14.8. This position is off the VLA coordinates of \citet{nagar05} by
-0.5\arcsec~in right ascension and -0.1\arcsec~in declination. Given
the $2\farcs6 \times 2\farcs3$ beam, this source is however still
identifiable with the radio source, despite the large offset in right
ascension. The 1 mm continuum
map shows a weak peak at the same location. Fitting a point-source
model in the $uv$-plane to the same coordinates as found for the 3 mm
continuum emission gives a flux density of 1.56 mJy with a rms noise
$\sigma=0.75$
mJy. We thus do not consider this a reliable determination of the flux density
at 1 mm, and instead adopt a $3\sigma$ upper limit of 2.25 mJy. 

\subsection[] {CO line emission}

\begin{figure*}
\begin{center}
  \rotatebox{270}{\includegraphics[width=7cm]{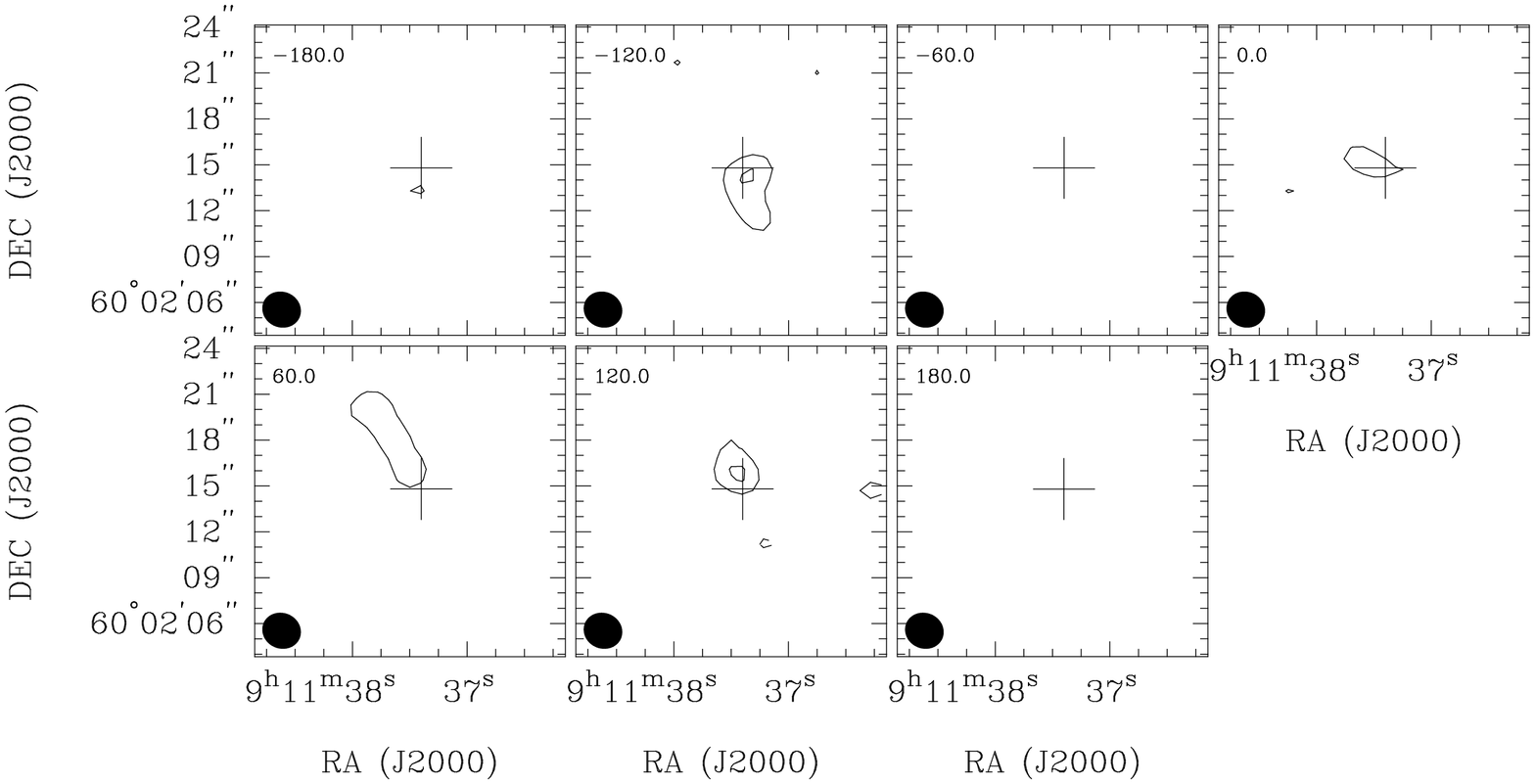}}
\rotatebox{270}{\includegraphics[width=7cm]{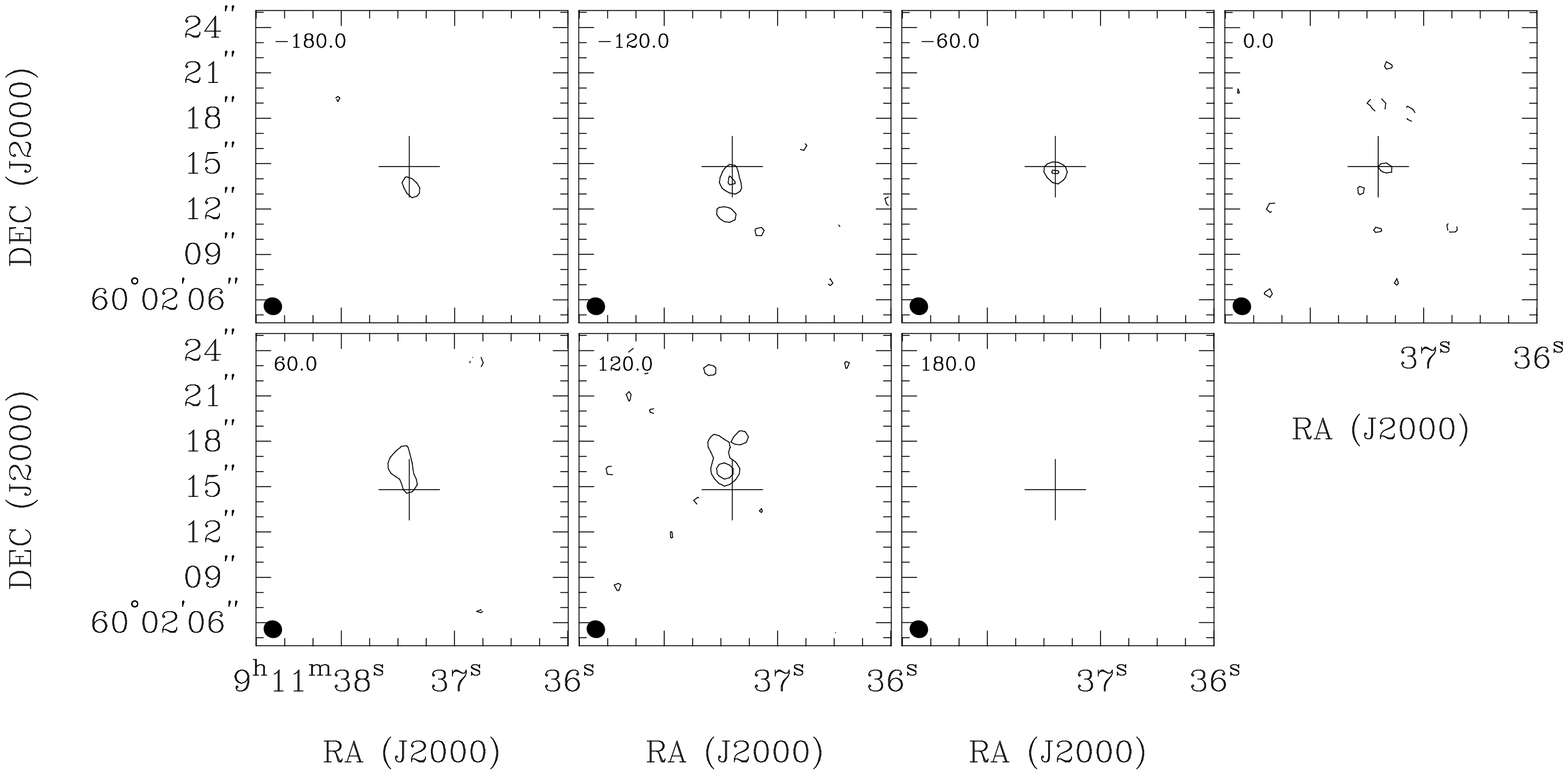}}
\caption{Channel maps of NGC 2768. {\em Top:} $^{12}$CO(1$-$0) line. {\em Bottom:}
  $^{12}$CO(2$-$1) line. The channels are
  60 km s$^{-1}$ wide and contours are plotted at -3 (dashed), 3, 6,
  and 9 $\sigma$
  with $\sigma =$ 1.0 mJy beam$^{-1}$ for CO(1-0) and $\sigma =$
  1.6 mJy beam$^{-1}$ for CO(2-1). The number in the top left
  corner of each frame is the central velocity of that frame, relative
  to the observed central velocity of $V_{\mathrm{sys}}=1373$ km s$^{-1}$. The
  cleaned beam is shown in the bottom left corner
  of each frame. The cross marks the centre of the 5~GHz radio
  continuum \citep{nagar05}.} 
\label{fig:chmap10}
\end{center}
\end{figure*}

 \begin{figure*}
\begin{center}
\rotatebox{0}{\includegraphics[width=7cm]
  {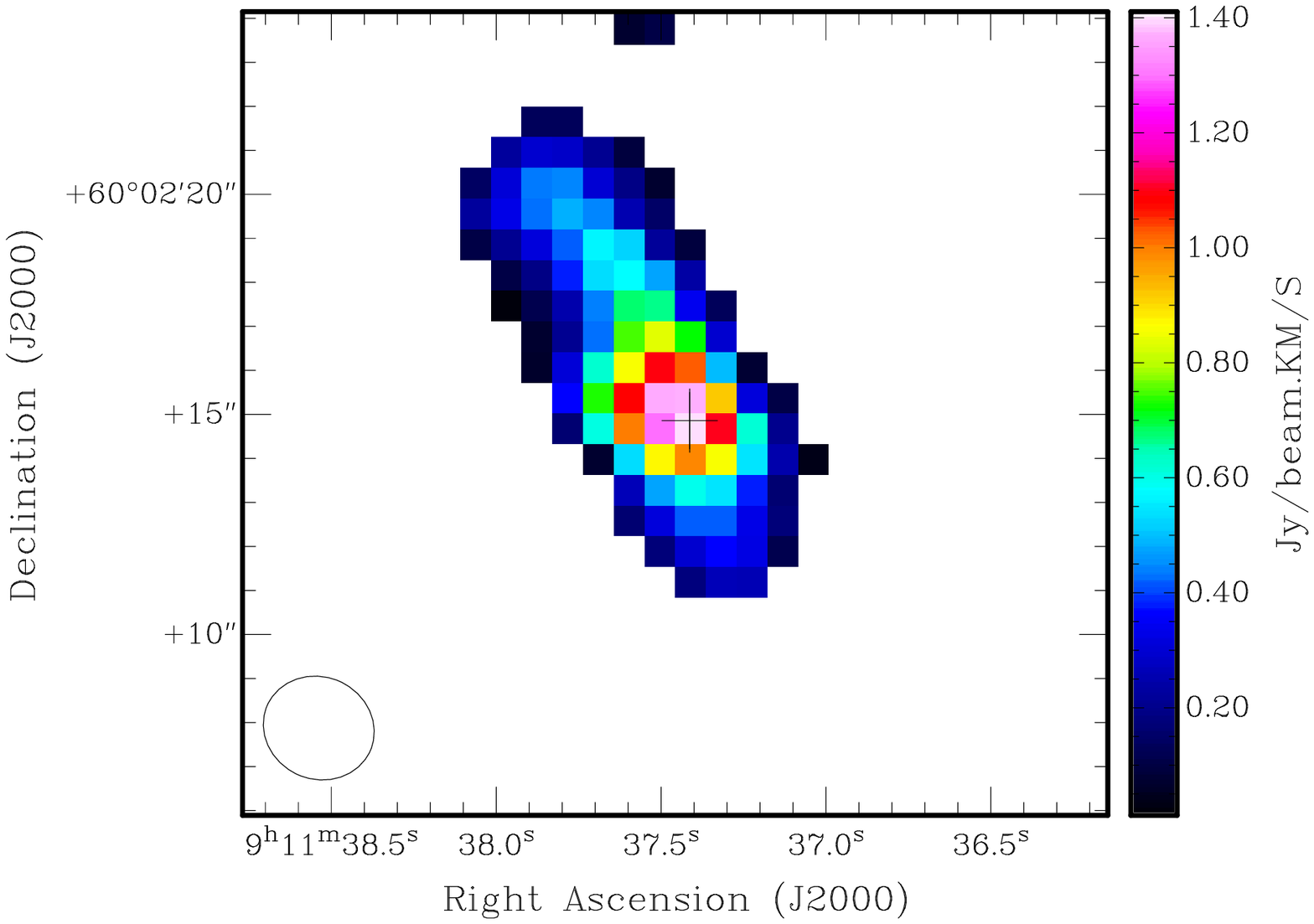}}
\rotatebox{0}{\includegraphics[width=7cm]
  {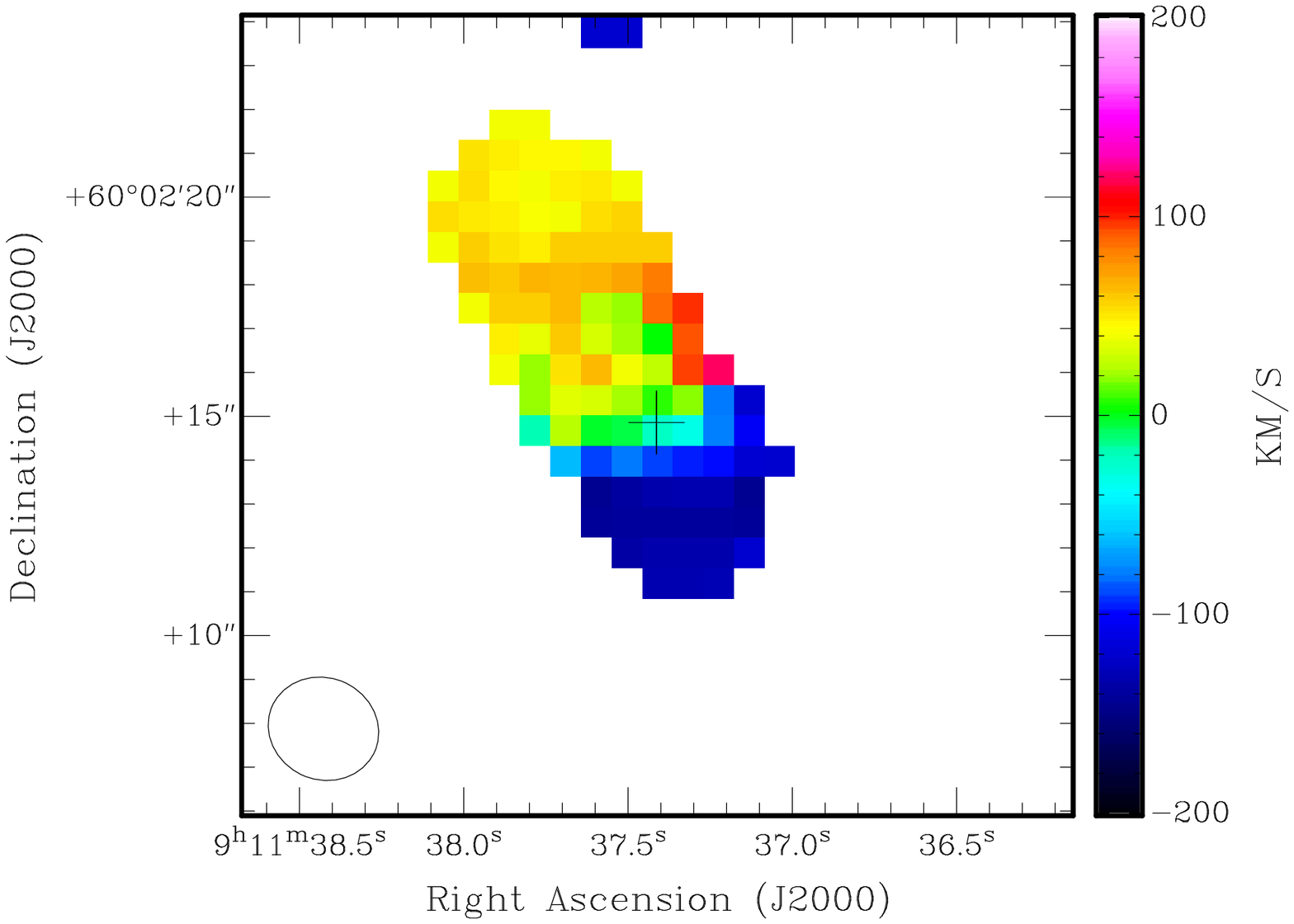}}\\
\rotatebox{0}{\includegraphics[width=7cm]
  {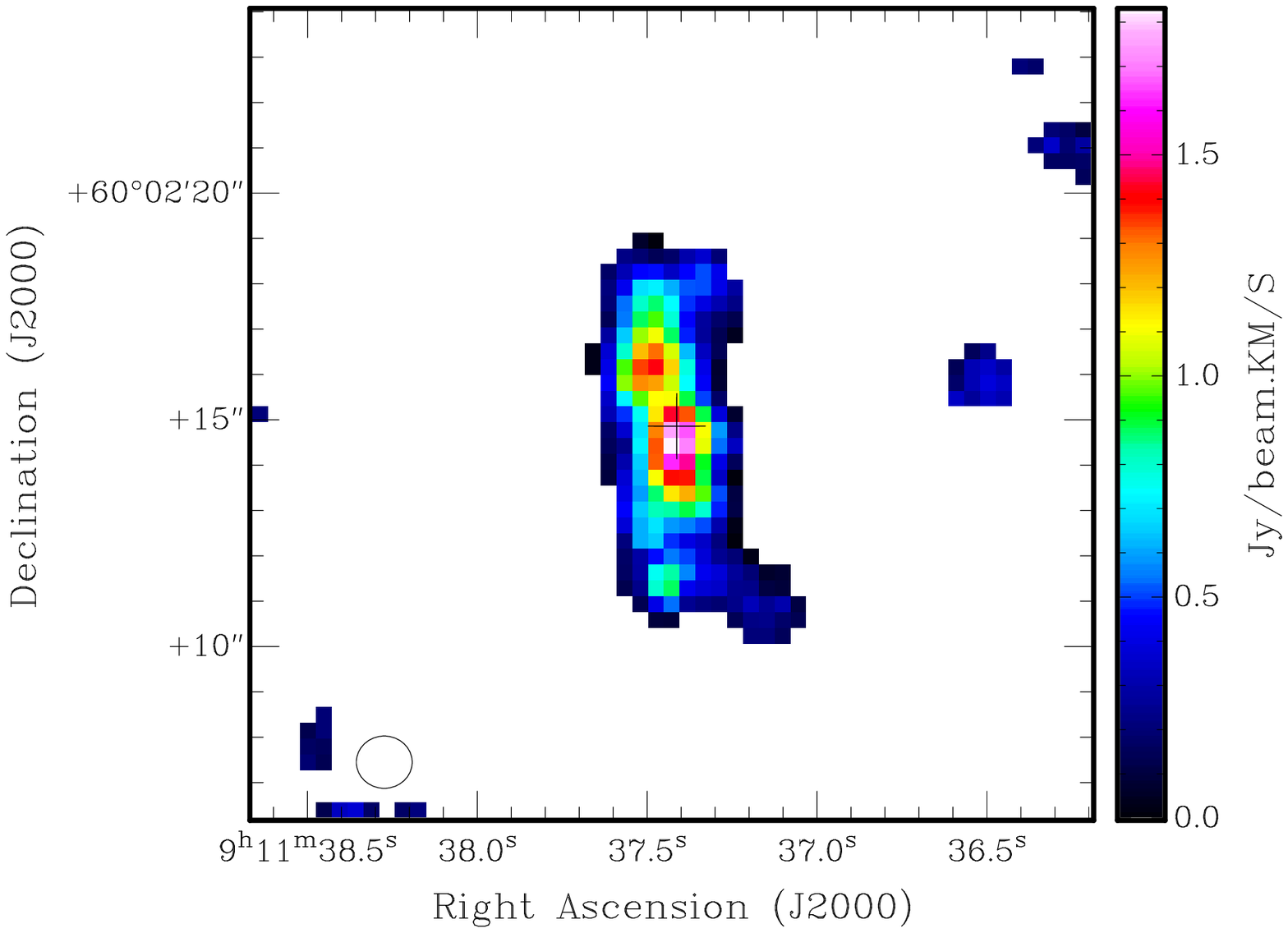}}
\rotatebox{0}{\includegraphics[width=7cm]
  {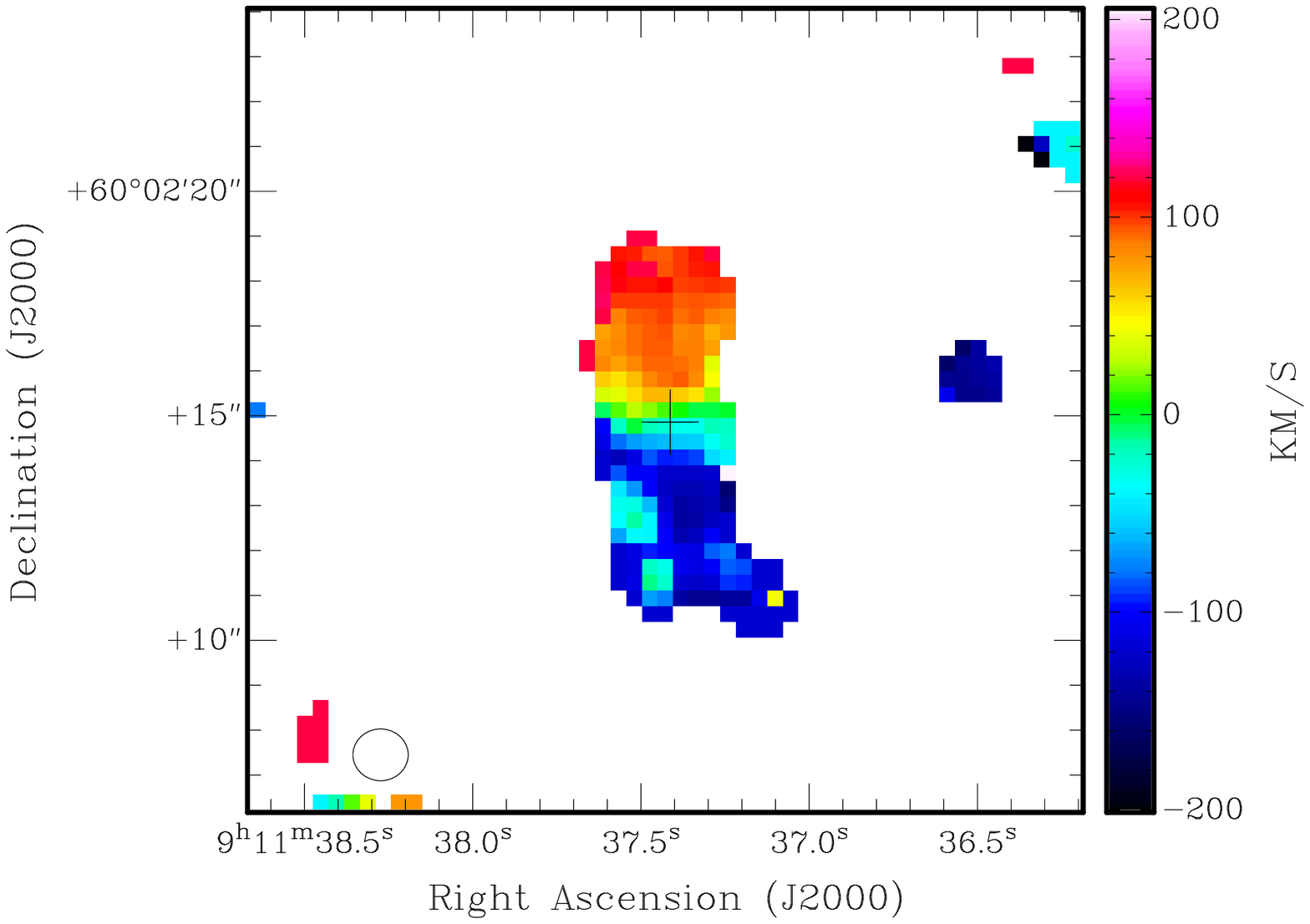}}
\caption{ {\em Left:} CO integrated intensity maps of NGC~2768. {\em
    Right:} CO mean velocity maps. Top: CO(1-0). Bottom: CO(2-1). The synthesized beam is shown in the
  bottom-left corner of each map. The black cross marks the centre of
    the 5~GHz radio continuum \citep{nagar05}. White pixels have been masked according to
    the procedure described in Section~3.2.}
\label{fig:meanvelo}
\end{center}
\end{figure*}

\begin{figure}
\begin{center}
\rotatebox{270}{\includegraphics[width=3.9cm]{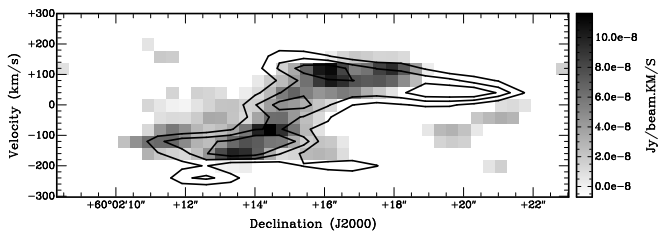}}
\caption{Position-velocity diagram of NGC~2768 along the polar axis. The pixels show the
  CO(2-1) emission, while the contours show the CO(1-0) emission at 3,
  6, and 9 $\times 10^{8}$~Jy beam$^{-1}$ km s$^{-1}$. White pixels have been masked according to
    the procedure described in Section~3.2.}
\label{fig:pv}
\end{center}
\end{figure}

 Fig.~\ref{fig:chmap10} shows the
 channel maps for the CO(1-0) and CO(2-1) lines from the 60 km
 s$^{-1}$ datacubes. To draw
 attention to the overall morphology and kinematics of the CO, we 
 created integrated intensity maps, mean
 velocity field maps and position-velocity (PV) diagrams using the 40 km s$^{-1}$ datacubes, for better
 velocity resolution (Figs.~\ref{fig:meanvelo} and \ref{fig:pv}). To make these maps,
 we first created a smoothed cube by smoothing with a 2D gaussian
 (FWHM of 4 pixels) spatially and hanning smoothing by 3 channels in
 velocity. We then computed the moments by integrating in velocity over the pixels
 in the original cube that corresponded to pixels above 3$\sigma$ in
 the smoothed cube. As the emission is oriented primarily north-south,
 we created the PV diagrams by computing the zeroth moment of the
 masked cube in the
 Declination-velocity plane. This smoothing-masking procedure excludes much,
 but not all, noise from the moment maps.  

 As shown in the integrated intensity maps of both transitions, the CO is distributed
somewhat asymmetrically, although with the signal just above 3$\sigma$
in most of the velocity channels, we have to be wary of the contribution of
noise even with the smoothing and masking procedure applied. The smaller
CO(2-1) beam resolves two peaks in the CO 
 intensity, one directly south of the galaxy centre and one to the 
 north by north-east. The two clear CO(2-1) peaks suggest that the
 densest gas might be in a disc or ring seen nearly edge-on. The
 lower-resolution CO(1-0) data blend these two separate peaks into one
 peak elongated in the direction of the CO(2-1) peaks. In
 addition, the CO(1-0) data pick up an extended structure along
 PA$\approx30^{\circ}$, especially prominent to the north-east. 
 While this feature is absent in the CO(2-1) maps, it appears in both
 the 40 and 80 km s$^{-1}$ channels of the CO(1-0) data, suggesting it
 is not just noise. The CO(2-1) observations must miss this molecular gas through
 some combination of the more limited primary beam size, noise, and a
 potentially lower excitation temperature of the infalling gas. 
Overall, the molecular gas distribution suggests that we
 are observing the formation of an inner polar ring or disc, with gas
 in the outskirts of the inner kiloparsec still being
 accreted onto the settled structure. 

 The velocity maps (Fig.~\ref{fig:meanvelo}) and PV
 diagrams (Fig.~\ref{fig:pv}) support this interpretation. The overlay
 of the CO(1-0) PV contours on the 
 CO(2-1) PV pixel map in Fig.~\ref{fig:pv} shows that the rotation of the main
 body of molecular gas is in good agreement for the two lines. 
 Unfortunately, the quality of the data does not allow us to
 reliably determine whether the molecular gas is in a ring or a
 disc. For simplicity and given the hint of a flattening of the
 rotation at offsets larger than $\approx22\arcsec$, we
 refer to it as a disc for the remainder of the paper but refrain from
 making any claims that depend on whether it is a disc or a ring. The polar rotation is
 consistent with the velocity field of the more extended ionized gas
 \citep{sarzi06}. However, the CO(1-0) mean  
 velocity map is irregular in the northern part of the galaxy. This
 irregularity is caused by the lower velocity (40 and 80 km s$^{-1}$) molecular gas making up the extension
 to the north-east of the galaxy centre, as can be clearly seen in the
 CO(1-0) PV contours. The velocity and extent of this
 gas lead to the conclusion that it must not yet be part of the
 central regularly rotating structure, but instead be in the process
 of accreting onto this structure. While other minor irregular
 features in the 
 velocity maps may indicate further irregularities in the molecular
 gas kinematics, the accidental inclusion of a noise peak from a
 particular channel map could influence the velocity map in the same
 way, thus we hesitate to analyse these features.

\begin{figure}
\begin{center}
\rotatebox{270}{\includegraphics[width=6cm]
{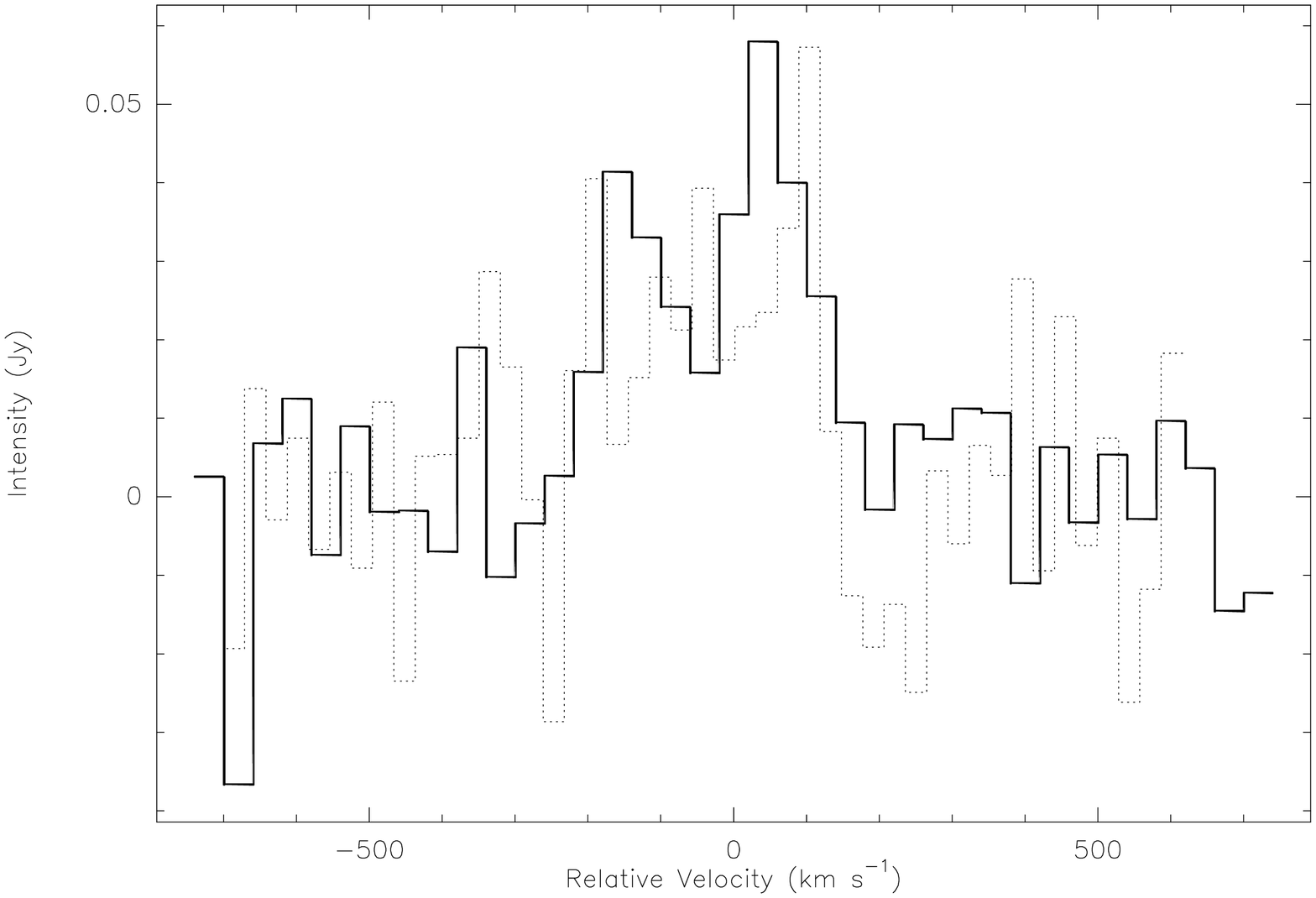}}
\vspace*{3mm}\\
\rotatebox{270}{\includegraphics[width=6cm]
{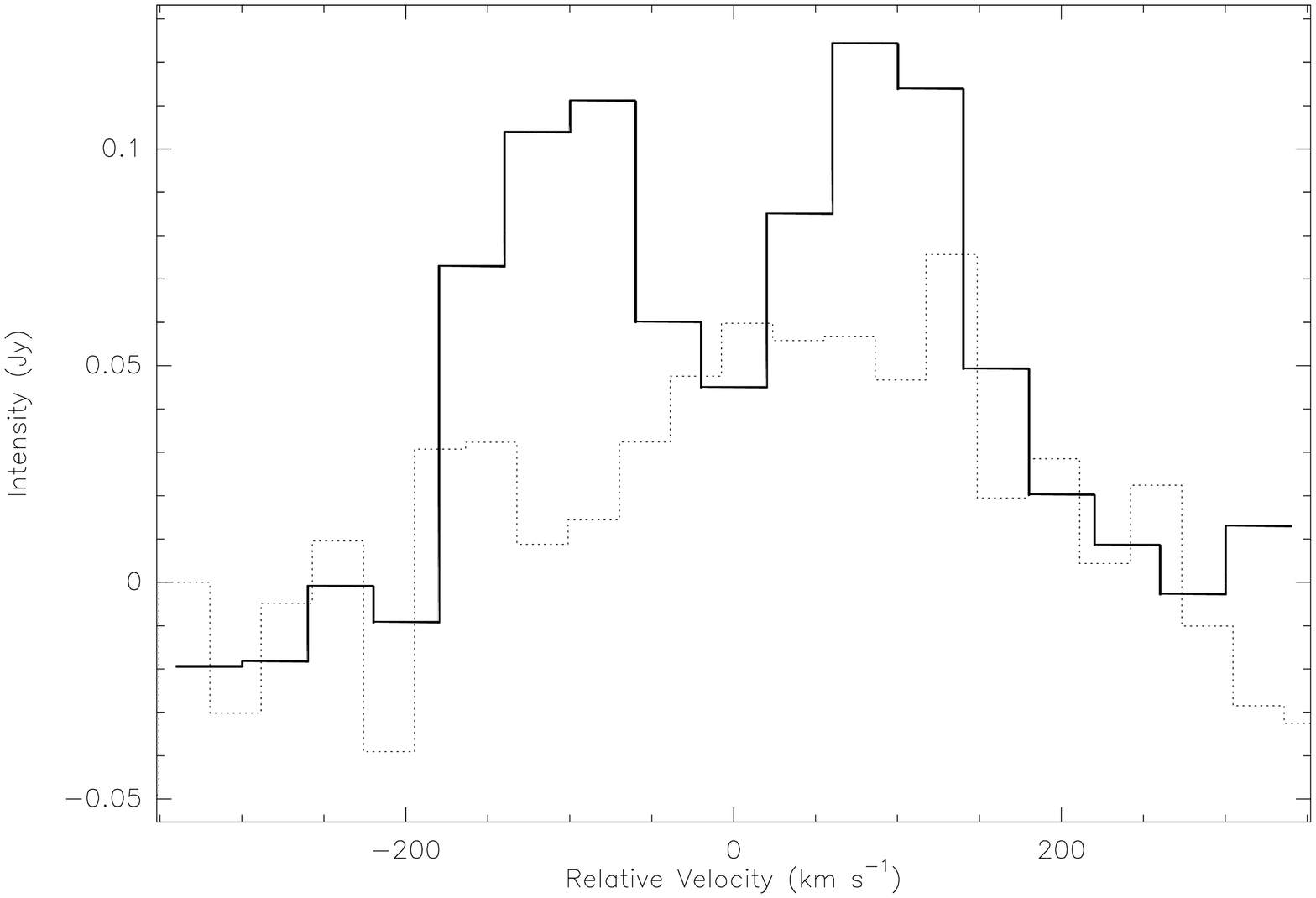}}
\caption{Solid lines show integrated spectra of the CO(1-0) line
  (top) and CO(2-1) line (bottom) in NGC~2768. The dotted lines show the
  single-dish spectra of \citet{CYB07}.}
\label{fig:spec}
\end{center}
\end{figure}

To construct integrated spectra and determine total fluxes, we
selected the pixels in the central non-zero parts of the two integrated
intensity maps as our spatial regions of interest. Integrating the
original datacubes over these spatial regions gave us the spectra shown
in Fig.~\ref{fig:spec}. Spectra from the 30m single-dish observations of
\citet{CYB07} are over-plotted in dotted lines.

Comparison with the \citet{CYB07} 30m data shows that the double-peaked
profile seen in both lines in the interferometric data is absent in
the 30m single-dish data. While the large amount of noise in the 30m
spectra certainly contributes to the difference in the spectral
profiles, we also note that both 30m spectra are biased toward the
positive relative velocities seen on the northern side of the galaxy. If
the 30m's pointing were off to the north, it would explain why
intensity is preferentially missing at the negative
velocities. However, the intensities at positive relative velocities
at 1 mm are also much lower than found with the interferometer. This
difference suggests that the pointing may have also been off in the
east-west direction. The 1 mm spectrum suffered much more from this
pointing error as the half-power beam-width is only 11\arcsec. 
 We note that errors in the flux calibrations could also explain some of
the discrepancy. 

\begin{table}
 \caption{CO fluxes and molecular mass estimates in NGC~2768.}
 \begin{tabular}{@{}lccc}
  \hline
  Line & Instrument & Flux & M$_{\mathrm{H_{2}}}$\\
   & & (Jy km s$^{-1}$) & ($10^{7}$~M$_{\sun}$)\\
  \hline
  $^{12}$CO(1-0) & 30m & $10.4\pm1.6\pm1.0$ &  $6.4\pm1.0\pm0.6$\\
  $^{12}$CO(1-0) & PdB & $11.3\pm1.3\pm2.3$ & $6.8\pm0.8\pm1.4$  \\
  $^{12}$CO(2-1) & 30m & $18.6\pm2.5\pm1.9$  \\
  $^{12}$CO(2-1) & PdB & $30.6\pm1.7\pm6.1$   \\
  \hline
 \end{tabular}
Errors reported are random and systematic (flux calibration) errors,
respectively. 
\label{tab:COFlux}
\end{table}

To estimate the total CO flux in each line, we integrated the 
spectra over the velocity range with observed emission, i.e. from 1193 to 1553 km s$^{-1}$. Table~\ref{tab:COFlux} reports the 
calculated total fluxes, 11.3 Jy km s$^{-1}$ at CO(1-0) and 30.6 Jy km
s$^{-1}$ at CO(2-1). Fluxes for the 30m observations of \citet{CYB07}
are also reported; these fluxes have been calculated over identical velocity
ranges as their interferometric counterparts. For CO(1-0), the
fluxes agree within the noise 
and the errors in the flux calibration. For the CO(2-1) line, the 30m
misses much of the flux seen by the interferometer, especially at
negative relative velocities where the flux from the 30m is 8.2 Jy km
s$^{-1}$ less than for the interferometer. As we have noted above, we
believe much of this difference to be due to a pointing error.

 We used the total fluxes in CO(1-0) to compute the total molecular
 hydrogen mass using the formula $M(\mathrm{H}_2) = (1.22 \times 10^4
 \mathrm{M}_{\sun})\;D^2*S_{\mathrm{CO}}$, where D is the distance measured in Mpc and
 $S_{\mathrm{CO}}$ is the total CO(1-0) flux. This formula
 comes from using the standard CO to H$_{2}$ conversion
 ratio $N(\mathrm{H}_2)/I(\mathrm{CO}) = 3 \times 10^{20}$ cm$^{-2}$, where $N(\mathrm{H}_{2})$ is
 the column density of H$_{2}$ and $I(\mathrm{CO})$ is the CO(1-0) intensity in
 K~km~s$^{-1}$. The total H$_{2}$ mass 
 detected by our interferometric observations is then
 $6.8\times10^{7}$ M$_{\sun}$. As nearly all the CO(1-0) flux 
 comes from
 the nascent polar disc seen in the integrated intensity maps, this
 gives a rough estimate of the molecular mass already in the central polar disc. 
We estimate the integrated line ratio $R_{\mathrm{21}} = $
 CO(2-1)/CO(1-0)$\approx0.7$, indicating optically thick molecular gas
 that is subthermally excited, as expected for a non-starburst galaxy. This ratio is computed assuming we have
 mapped the full extent of the CO in both lines in our interferometric
 maps. As the CO
 distribution in both lines is only about half the diameter of the HPBW of
 the primary beam, this assumption is reasonable.

\section{Discussion}

\subsection{Origin of the Molecular Gas}

\begin{figure*}
\begin{center}
\includegraphics[width=16cm]
{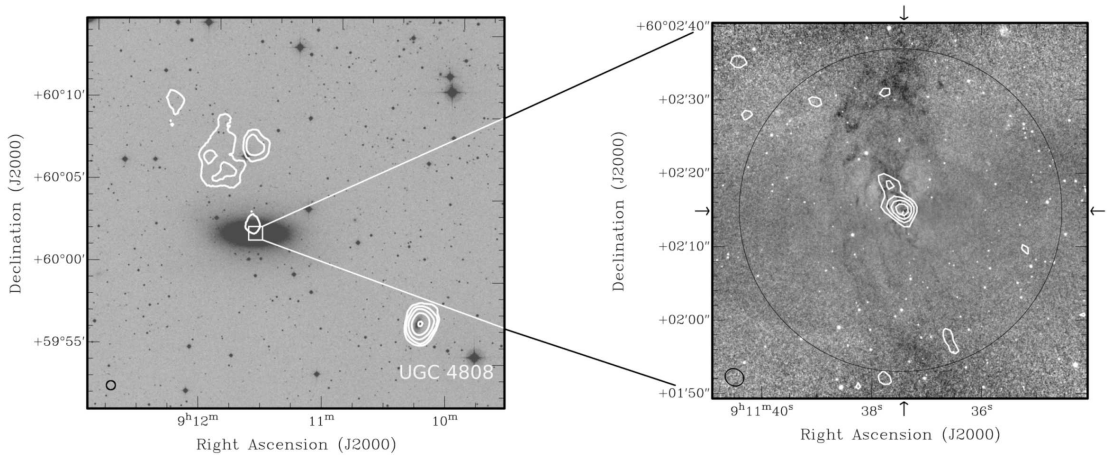}
\caption{The cool ISM in NGC~2768. The left panel shows the H$\:${\small I} surface density
  distribution in contours of 2, 4, 8, 16 and $32 \times 10^{19}$~atom~cm$^{-2}$
  plotted over an R-band Digital Sky Survey (DSS) image. UGC~4808 is
  visible in the bottom right-hand corner. The right
  panel shows the CO(1-0) intensity distribution in contours of 0.5,
  0.75, 1.0 and 1.25 Jy~km~s$^{-1}$ plotted over a dust map of
  NGC~2768 obtained by dividing the HST F814W ACS image by an ellipse
  model fit. In the right panel, the outer circle indicates the 44\arcsec~
  HPBW of the CO(1-0) primary beam of the interferometer and the
  arrows indicate the galaxy centre. The black ellipses in the bottom
  left corners of both panels show the synthesized beams of the H$\:${\small
  I} and CO observations, respectively.}
\label{fig:HIdust}
\end{center}
\end{figure*}

With H$\:${\small I} data from \citet{morganti06}, dust maps from HST images
(e.g. Fig.~\ref{fig:HIdust}),
and the molecular gas maps of section 3, we
now have a fairly complete census of the cool ISM in and around NGC~2768. The
most striking feature, besides the fact that the ISM lies in the polar
plane, is the clear link between all components. All observed phases
of the cool ISM extend to the north-east, first in CO, then dust,
then H$\:${\small I} (Fig.~\ref{fig:HIdust}).  This shared asymmetry 
ranges from the inner kiloparsec in the molecular gas to over 30 
kpc in H$\:${\small I}. This link between the
different ISM
components clearly indicates that they were all accreted from the same
source. It also suggests that the gas accretion will continue for some
time, perhaps in discrete episodes.

This extended and asymmetric morphology along with the misaligned kinematics of
the cool gas also firmly indicates an external origin. We consider three
options: a minor merger with a gas-rich
dwarf, tidal accretion from a gas-rich galaxy, and accretion from an H$\:${\small I}
filament. The regularity of both the photometry and stellar kinematics of
NGC~2768 argues against the recent merger of a gas-rich dwarf galaxy.
No stellar tidal tail nor other irregularity has been observed, although the
data we have access to do not pose particularly stringent upper limits
(i.e. the optical photometry is relatively shallow).
The accretion of primordial (metal-free) H$\:${\small I} scenario fails to explain
both the dust and CO observed. We
thus consider tidal accretion as the most likely scenario. 

\begin{table*}
 \caption{Basic properties of galaxies near NGC~2768. This table lists
 all galaxies with a projected distance differing by less than
 400~kpc and a redshift differing by less than 400 km s$^{-1}$
 from NGC~2768. Only galaxies with known redshifts are included.}
 \begin{tabular}{@{}lrrrrrrrc}
  \hline
  Name & $\Delta\theta$ & $\Delta$R & $\Delta$V$_{\mathrm{sys}}$ & $B_{{\mathrm
  T}}$ &
  $M_{{\mathrm{ H\:I}}}$ & $M_{{\mathrm{ H\:I}}}$/$L_{B}$ & $\Delta t$ & Type\\
   & & (kpc) & (km s$^{-1}$) & (mag) & ($10^{8}$ M$_{\sun}$) &
  (M$_{\sun}$/L$_{\sun}$) & (Gyr) & (NED)\\ 
  (1) & (2) & (3) & (4) & (5) & (6) & (7) & (8) & (9) \\
  \hline
NGC~2768 & -- & -- & -- & $10.60$ & $1.7$ & $0.007$ & -- & E\\
UGC~04808 & $11\farcm5$ & $72$ & $-78$ & $15.03$ & $1.6$ & $0.41\phantom{0}$ &
  $0.38$ & Scd\\
PGC~025982 & $16\farcm5$ & $103$ & $-236$ & $14.71$ & &  & &\\
MAILYAN~038 & $19\farcm8$ & $124$ & $187$ & & $0.3$ &  & 1.71 & \\
PGC~2599651 & $21\farcm8$ & $136$ & $197$ & $17.05$ & $0.4$ & $0.63\phantom{0}$
  & 3.83 & \\
PGC~2601563 & $22\farcm5$ & $141$ & $-325$ & $17.36$ & $0.6$ & $1.4\phantom{00}$
  &  &\\
SDSS~J091531.98+594948.5 & $31\farcm9$ & $200$ & $-5$ & $17.14$ &  &
 &
0.98 & \\
NGC~2742 & $40\farcm2$ & $252$ & $-84$ & $11.38$ & $18.4$ & $0.16\phantom{0}$ &
1.36 & SA(s)c \\
NGC~2726 & $50\farcm6$ & $317$ & $145$ & $12.56$ & $1.8$ & $0.05\phantom{0}$ &
2.25 & Sa? \\
  \hline
 \end{tabular}

 (1) Galaxy name. (2) Angular separation. (3) Projected separation
 (assuming a distance of 21.5 Mpc). (4) Relative line-of-sight velocity. (5)
 Total corrected apparent blue magnitude (HyperLEDA). (6) Mass of
 neutral hydrogen (Morganti et al. 2006, except for NGC~2742 --
 Broeils \& van Woerden 1994 and NGC~2726 -- Haynes et al. 1988). (7)
 H$\:${\small I} mass to total blue luminosity
 ratio. (8) Estimate for time since a possible interaction with NGC
 2768. (9) Galaxy classification (NED).  
\label{tab:neighbors}
\end{table*}

The loose-group centred on NGC~2768, catalogued as Lyon Group of
Galaxies (LGG) 167 \citep{garcia93}, 
contains many potential gas-donor galaxies (see
Table~\ref{tab:neighbors}). The H$\:${\small I} datacube obtained
by \citet{morganti06} reveals neutral hydrogen around $4$ lower-luminosity
galaxies within a 150 kpc projected separation from NGC~2768. The H$\:${\small I} velocities
indicate group membership for all of
these galaxies. Considering the wide range (0.1 to 6) of H$\:${\small
  I} mass to blue luminosity ratio ($M_{{\mathrm{ H}\:I}}$/$L_{B}$) values found in late-type dwarf galaxies
\citep{swaters99}, any of these galaxies could have provided the
$\sim10^{8}$ M$_{\sun}$ of H$\:${\small I} found around NGC~2768 without requiring
an unreasonable amount of H$\:${\small I} for its blue luminosity.

Including all galaxies within 400 kpc in projected separation and 400
km s$^{-1}$ in relative line-of-sight velocity from NGC 
2768, we created a list of known galaxies that may have
recently interacted with NGC~2768 (Table~\ref{tab:neighbors}). Given the
relaxed appearance of the ionised gas in the SAURON map
\citep{sarzi06}, we estimate that this gas must have been present for
at least 3-10 dynamical timescales, or 0.2-0.7 Gyr. Assuming $\sigma_{\mathrm{group}}=200$ km s$^{-1}$
as a reasonable three-dimensional velocity dispersion for a small
group, we estimated the time since each galaxy could have interacted
with NGC~2768,
\[
\Delta t = \frac{\Delta R}{\sqrt{\sigma_{\mathrm{group}}^2 - \Delta V_{\mathrm{sys}}^2}},
\] where $\Delta R$ is the projected separation and $\Delta
V_{\mathrm{sys}}$ the line-of-sight velocity relative to NGC~2768.
Values of this estimated interaction time are listed in
Table~\ref{tab:neighbors}. This analysis shows that the nearby Scd
galaxy UGC~4808 is most likely to have interacted with NGC~2768
recently and provided the cool ISM. UGC~4808 can be seen in the bottom
right corner of Fig.~\ref{fig:HIdust}. Its alignment with the
large-scale H$\:${\small I} supports this conclusion.

\subsection{Fate of the Molecular Gas: Star Formation?}

In NGC~2768, we observe no definitive signs of young stars as seen in
 the other five SAURON early-type galaxies with mapped CO emission
 \citep*[NGC~2865, Schinnerer \& Scoville 2002; NGC~3032, NGC~4150,
 NGC~4459 and NGC~4526,][]{young08}. Excluding NGC~2685, in which the CO
 lies outside of the SAURON field-of-view, young stars in these galaxies are
 traced by high H$\beta$ linestrengths (all), cold stellar components that
 share the CO kinematics (all except NGC~3032), and low [O$\:${\small
 III}] to H$\beta$ emission line ratios (all except NGC~4150).

\begin{figure*}
\begin{center}
\rotatebox{0}{\includegraphics[width=7cm]
  {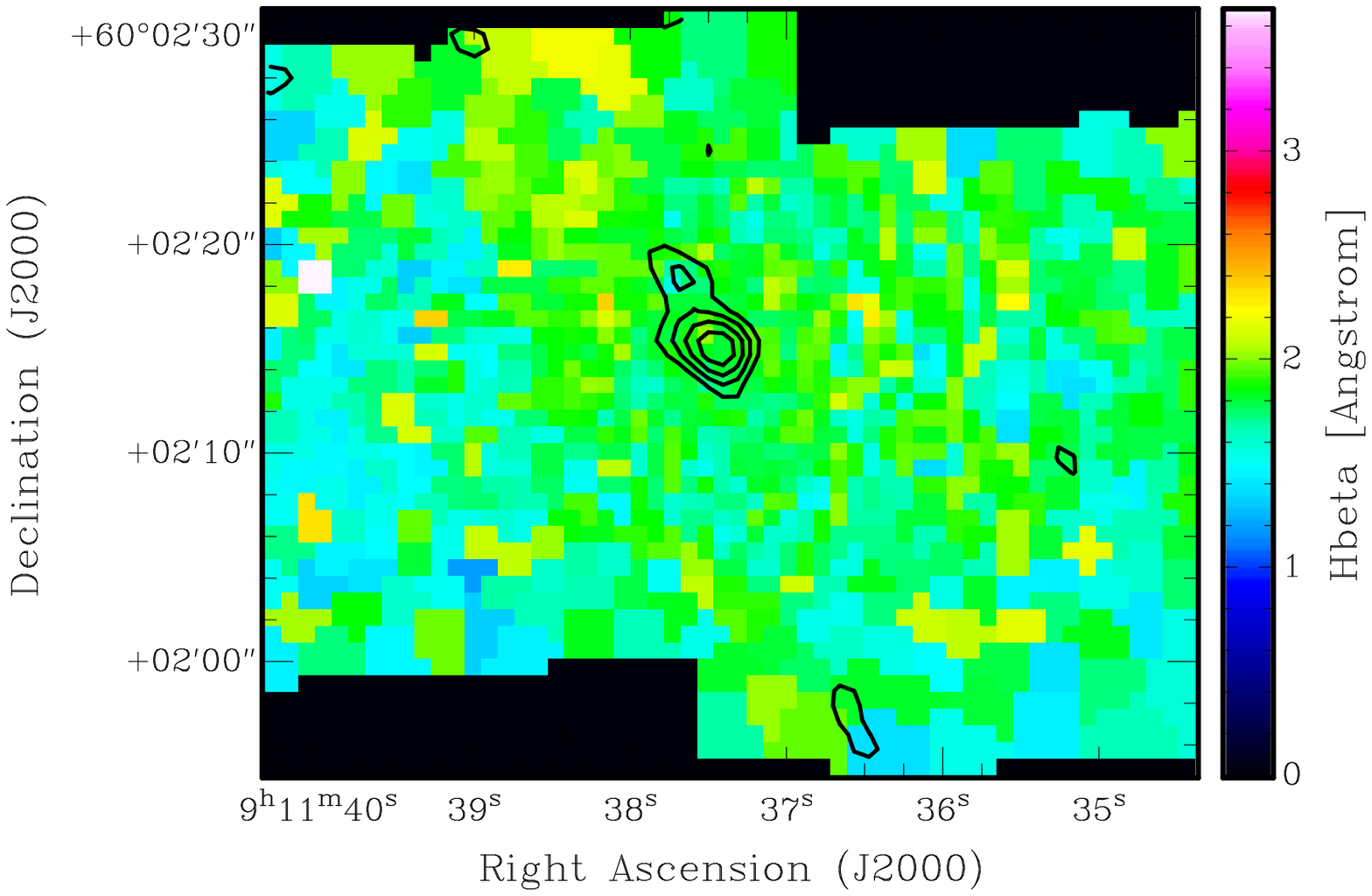}}  
\rotatebox{0}{\includegraphics[width=7cm]
  {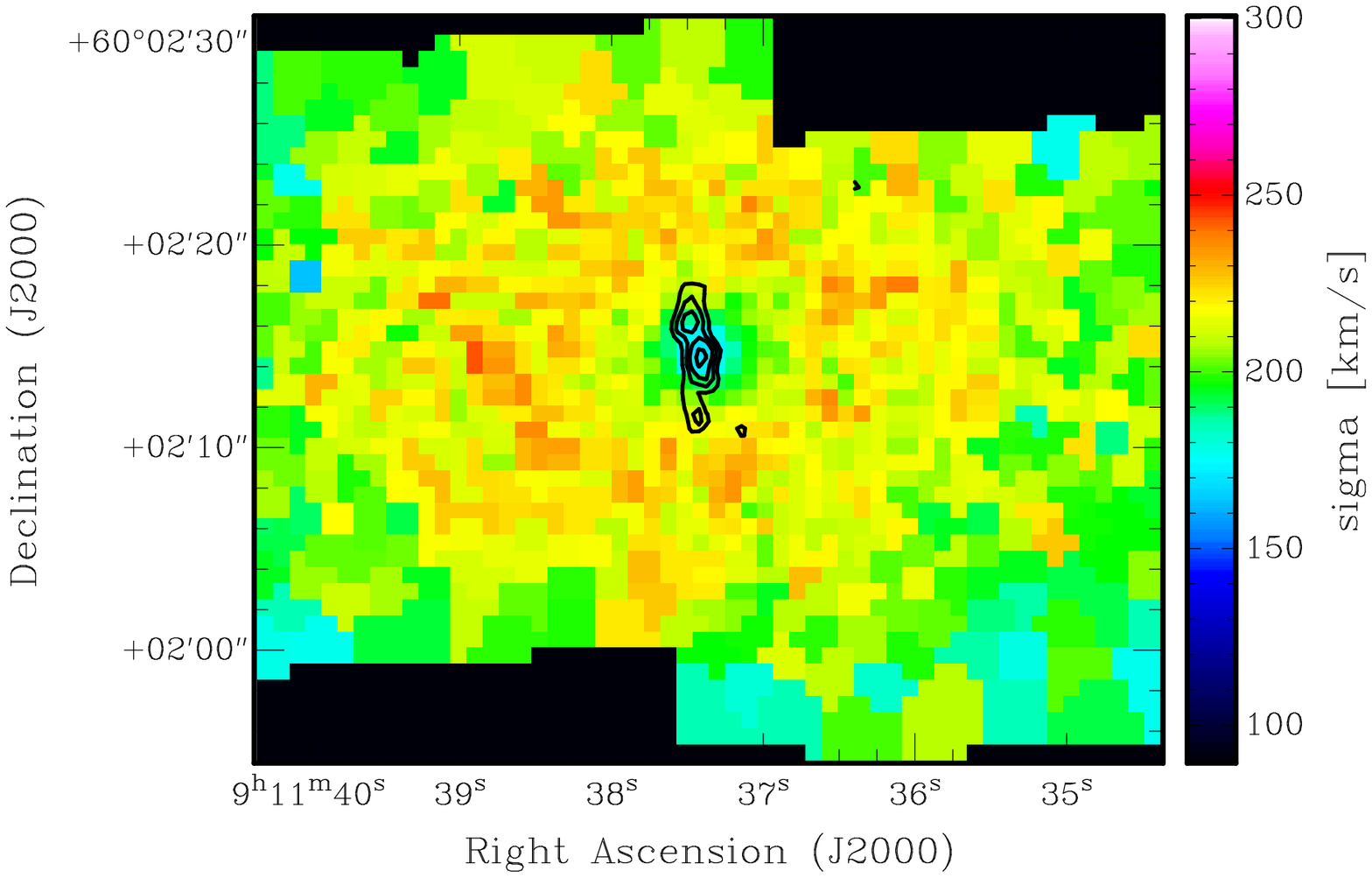}}\\
 \rotatebox{0}{\includegraphics[width=7cm]
  {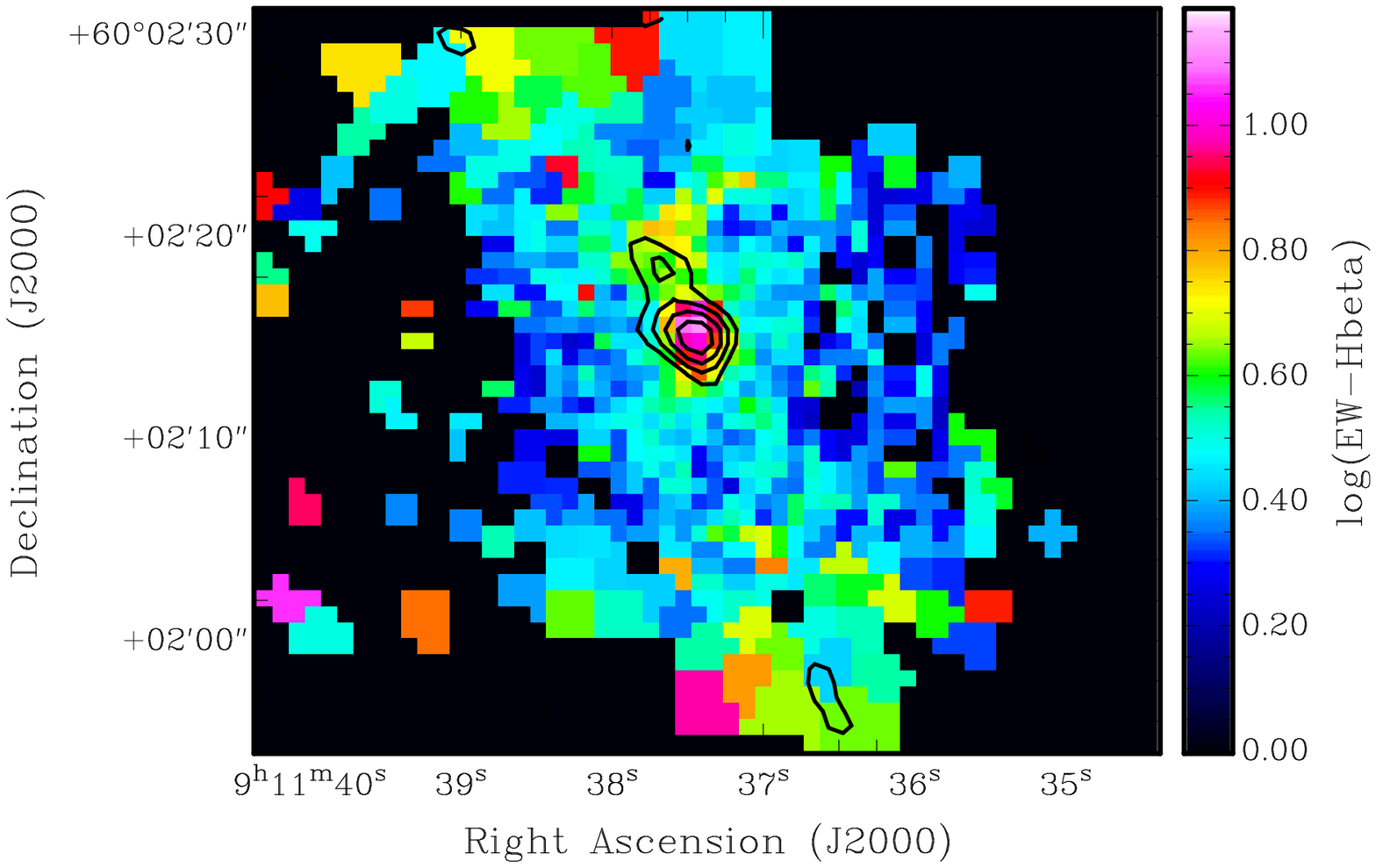}} 
\rotatebox{0}{\includegraphics[width=7cm]
  {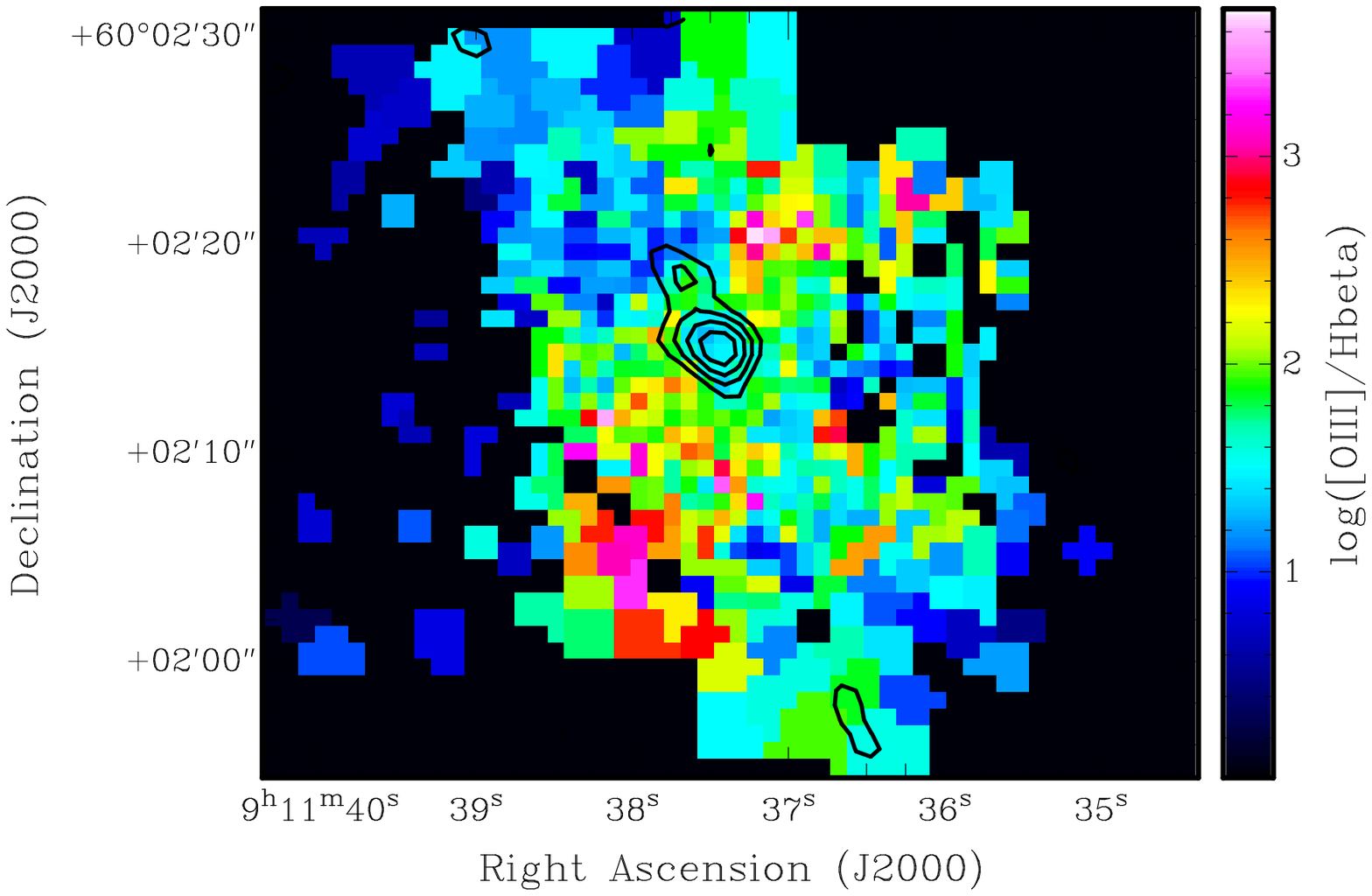}}
\caption{{\em Top Left:} CO(1-0) contours of NGC~2768 over the SAURON H$\beta$
  absorption linestrength map \citep{kuetal06}. No significant change in linestrength is coincident with the
  molecular gas. {\em Top Right:} CO(2-1) contours over the stellar
  velocity dispersion map \citep{emsellem04}. The CO is co-spatial with
  the dispersion drop. {\em Bottom Left:} CO(1-0) contours over a 
  map of the equivalent width of H$\beta$ emission \citep{sarzi06}. 
  Note the increase in equivalent width near the extended CO peak to 
  the north-east. 
  {\em Bottom Right:} CO(1-0) contours over a map of the flux ratio
  of [O$\:${\small III}] to H$\beta$ emission \citep{sarzi06}.}
\label{fig:starform}
\end{center}
\end{figure*}

The H$\beta$ linestrength index of NGC~2768  is
 1.70 \AA\ within a central 9\arcsec~radius aperture (one eighth of
 the effective radius) and  
remains near this level over the entire SAURON 
field (Kuntschner et al. 2006; Fig.~\ref{fig:starform}). Higher resolution but lower
signal-to-noise data from the OASIS integral-field spectrograph give
slightly larger values for the H$\beta$ linestrength
\citep{mcdermid06}. However, the discrepancy between the OASIS and SAURON 
values is probably attributable to the difficulty of subtracting the
 strong H$\beta$ 
emission before fitting the linestrengths, and the higher signal-to-noise data 
from SAURON provide a better estimate. Combining line-strength indices
 (H$\beta$, Fe5015, and Mg {\it b}) with stellar population models,
 Kuntschner et al. (in preparation) show that NGC~2768 has a
 fairly uniform, old stellar population. However, linestrengths
 are not sensitive to very recent star formation nor to a very small
 percentage of young stars superposed on a dominant older stellar
 population, so linestrengths cannot rule out particularly recent and/or
 weak star formation.

The SAURON map of the stellar velocity dispersion shows a dramatic drop of 
$\approx45$ km s$^{-1}$ in the
innermost 8$\arcsec$ (Emsellem et al. 2004; Fig.~\ref{fig:starform}). This drop in
dispersion is also clearly seen in long-slit data \citep{simien} and in
OASIS data \citep{mcdermid06}. 
Young stars born from dynamically cold gas in a cicumnuclear disc
can produce such a drop in dispersion \citep{emsellem01, wozniak03}, but
they should also produce a noticeable effect in the velocity map. 
However, the velocity field of NGC~2768 lacks any signature of a
dynamically cold stellar disc or ring. Thus the stellar velocity dispersion drop cannot be
directly attributed to the formation of a stellar component either
from the molecular polar ring itself, or from gas settled into the
main plane of the galaxy as may occur during polar accretion events
\citep{bournaud03}. Additionally, it is difficult to associate this
 drop to a young stellar population, as no coincident increase
in H$\beta$ linestrength is observed. We conclude that
no kinematic component visible in the SAURON maps is directly related to the molecular gas.

Particularly low ratios of [O$\:${\small III}] to H$\beta$
[log([O$\:${\small III}]/H$\beta$) $< -0.2$] indicate
star formation as the source of ionisation
\citep{ho97}. In NGC~2768, log([O$\:${\small III}]/H$\beta$) ranges
from about 0.0 to 0.5. In this range, the ionisation can come from
sources other than star formation. The four additional possibilities are heating by an
X-ray halo, an AGN, shocks or post-AGB stars. NGC~2768 lacks an X-ray
halo \citep{komossa99}, so this option is ruled out. The H$\beta$ flux
in NGC~2768 does not drop off with the square of the radius
as it should if ionised solely by an AGN \citep{hatch}, therefore everything
outside the central 2\arcsec~ must be ionised by another
source (Sarzi et al., in preparation).

 The most prominent feature of the ionised gas
in NGC~2768 is a roughly spiral-shaped region of increased
[O$\:${\small III}] equivalent width \citep[see fig. 4b,][]{sarzi06}. This strong enhancement in
[O$\:${\small III}] without a corresponding strong enhancement in H$\beta$ is
seen in other SAURON galaxies (NGC~2974, NGC~3414) and must be caused
by shock ionisation of the gas, as post-AGB stars are too
symmetric to cause such asymmetric ionisation and star formation would
not cause such an increase in [O$\:${\small III}]. Yet the smooth and circular H$\beta$ emission
suggests that an ionisation source other than shocks is responsible for the
ionisation outside of the spiral pattern. Both star formation
and post-AGB stars remain options. Post-AGB stars must be present in
such an old galaxy and thus must play a role. However, the H$\beta$
equivalent width map from SAURON shows a weak enhancement along the
extended molecular gas seen to the northeast of the galaxy
centre (see Fig.~\ref{fig:starform}), pointing to star formation in this ionised gas. 

The FIR luminosity (L$_{\mathrm{FIR}}$) to molecular mass (M$_{\mathrm{H}_{2}}$) ratio also
provides possible evidence of star formation \citep{CYB07}. We computed a bolometric
FIR flux for NGC~2768, 
\[ FIR \equiv 1.26 \times 10^{-14}\: \mathrm{W\: m^{-2}} (2.58 \;
S_{\mathrm{60\mu m}} + S_{\mathrm{100\mu m}}),\]
where $S_{\mathrm{60\mu m}}$ and $S_{\mathrm{100\mu m}}$ are IRAS fluxes
    in Jy. The total FIR luminosity is then L$_{\mathrm FIR} =
    (FIR)\; 4\pi D^{2} = 4.3\times10^{8}$ L$_{\sun}$, giving a
    L$_{\mathrm{FIR}}$ to M$_{\mathrm{H}_{2}}$ ratio of 6.3
    L$_{\sun}$/M$_{\sun}$, a reasonable value for a
    star-forming spiral \citep{devereux97}. In spiral galaxies, FIR emission
    traces star formation activity and the ratio of
    L$_{\mathrm{FIR}}$ to M$_{\mathrm{H}_{2}}$ is assumed
    to measure how efficiently stars are being formed out of the
    molecular hydrogen. If star formation is the dominant process
    contributing to FIR emission in NGC~2768, then star formation is
    proceeding at a normal efficiency. However, other sources (such as
    an AGN) may significantly contribute to the FIR emission in
    early-type galaxies, weakening this conclusion. We can still use the FIR luminosity
    to give an upper limit to the star formation rate, deriving an
    upper limit of 0.07 solar masses per year
    \citep[][eqn. 4]{kennicutt98}. Thus if star formation is present
    in NGC~2768, it is occurring at a very low rate, and the timescale
    to exhaust the supply of molecular gas is around 1~Gyr. 

As we have only unclear evidence of ongoing star formation in NGC
2768, we consider 
whether the molecular gas disc might be gravitationally stable and thus
unable to form stars. Theoretical models from \citet{kawata07} suggest
that circumnuclear discs in massive galaxies are more stable than
those in less massive galaxies. A stability analysis of the molecular
circumnuclear disc in 3C~31 based on Toomre's $Q$-parameter \citep{toomre64} suggests
that it is gravitationally stable and thus unable to form stars \citep{okuda05}.  Our data are insufficient to calculate
Toomre's $Q$ parameter by themselves, but comparing with the models of
\citet[][]{kawata07} (see their fig. 5), we find that the 0.5 kpc radius and
$6.8\times10^{7}$ M$_{\sun}$ mass of the molecular disc in NGC~2768
should be unstable, with a $Q$ parameter less than 1. Thus it is
reasonable to expect star formation in the polar disc of NGC~2768.

\subsection{Origin of the mm continuum}

From radio and X-ray observations, we know that NGC~2768 is a 
low-luminosity AGN. Very Large Array (VLA) and Very Long Baseline Array 
(VLBA) observations indicate a compact radio source with
no evidence of jets \citep{nagar05}. Most if not all 
X-ray emission is consistent with a point-source origin, and the X-ray count rate
differed between observations a month apart, indicating that the X-ray source
is variable \citep{komossa99}. 

\begin{figure}
\begin{center}
\rotatebox{270}{\includegraphics[width=7cm]{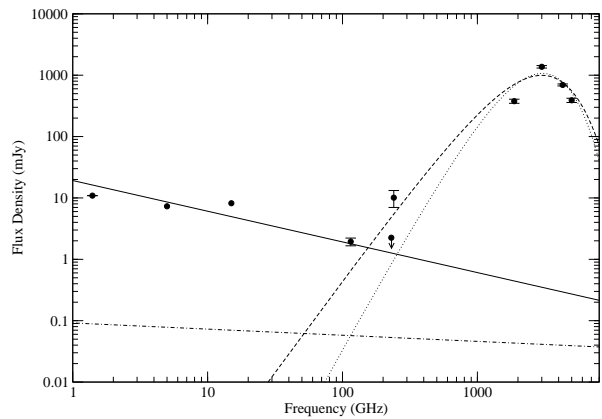}}
\end{center}
\caption{Spectral energy distribution of NGC~2768 from the radio to
  the far infrared (FIR). The solid line is an $\alpha=-0.5$
  power law fit,
  modelling synchrotron emission. The dot-dashed line estimates the
  maximum 
  contribution from free-free emission. The dashed curve extrapolates
  from the FIR dust bump to mm wavelengths using a $\beta=1$
  emissivity law; the dotted curves does the same with $\beta=2$.}
\label{fig:SED}
\end{figure}

The few LLAGN that have been studied at millimetre frequencies
show a range of spectral shapes attributed to varying contributions
from optically thin synchrotron emission, self-absorbed synchrotron
emission, and free-free emission
\citep[see ][]{nagar02,doi05,krips07}.
In Fig.~\ref{fig:SED}, we plot the radio through FIR
SED of NGC~2768, which is similar to those of some of the other LLAGN
previously observed. Values and references are listed in
Table~\ref{tab:SED}. We note that the flux densities at different frequencies
are necessarily at different resolutions and thus are not
entirely comparable. However, as no extended structures are
observed at the milliarcsecond resolution of the VLBA at 5 GHz, and the
1.4, 15, and 115 GHz detections are all consistent with a point-source
origin, we assume all these flux densities are dominated by an
unresolved source and 
can thus be compared without correction. The upper-limit at 230 GHz is
also for a point source at the phase centre at the resolution of
$1\farcs2\times1\farcs2$, while the 240 GHz detection is a bolometric
measurement from an 11\arcsec~beam, which could include a contribution
from cold dust \citep{wiklind95} as well as from a central AGN. These
two observations are thus not
as contradictory as they appear in the SED plot. However, we note
that our map of 1 mm continuum emission does not trace the dust pattern
seen in Fig.~\ref{fig:HIdust} nor do we see evidence of more emission in
the central 11\arcsec~after smoothing our data to this resolution.   

First, the lack of strong, coincident 1 mm continuum emission rules out
cold dust as the
origin of the 3 mm continuum emission, as the thermal spectrum from
dust drops sharply towards mm wavelengths (see the dust curves in Fig.~\ref{fig:SED}). We can also rule out 
free-free emission using the formula in \citet{doi05} which assumes an
electron temperature of $10^{4}$ K and
N(He$^{+}$)/N(H$^{+}$)$=0.08$:
\[\left(\frac{S_{\nu}}{\mathrm{mJy}}\right)\approx3.57\times10^{12}\left(\frac{F(\mathrm{H}\beta)}{\mathrm{erg\:cm^{-2}\:s^{-1}}}\right)\left(\frac{\nu}{\mathrm{GHz}}\right)^{-0.1}.\]
We use the total flux in H$\beta$ from the 
SAURON data \citep{sarzi06}, which gives an overestimate for the central
radio emission, as most of the H$\beta$ emission comes from ionising
sources other than star formation and only the flux in the central pixels should
correspond to H$\:${\small II} regions that could appear as a central
radio point source. Nevertheless, Fig.~\ref{fig:SED} reveals
that, even with this overestimate,  free-free emission is far too
faint to account for the 3 mm continuum.  

Having ruled out dust and free-free emission, processes directly
related with an accreting supermassive black hole are likely to be the source
of the 3 mm continuum detected in NGC~2768. Optically thin
synchrotron emission from compact jets or other ejecta will produce a
spectrum with a spectral index of $\alpha\leq-0.5$. The logarithmic
least-squares fit of an 
$\alpha=-0.5$ power law is shown in Fig.~\ref{fig:SED}. It roughly
fits the 3 radio data points and the 3 mm continuum emission, but the
low value at 1.4~GHz and the high value at 15~GHz hint at a flatter or
inverted spectrum, suggesting synchrotron self-absorption plays a role.

\begin{table}
 \caption{Spectral energy distribution of NGC~2768.}
 \begin{tabular}{@{}lccc}
  \hline
  Instrument & Frequency (GHz) & Flux Density (mJy) & Ref. \\
  \hline
  WRST & 1.4 & 10.9 & 1 \\
  VLBA & 5 & 7.3 & 2 \\
  VLA & 15 & 8.2 & 2 \\
  PdB & 115 & 1.94 & 3 \\
  PdB & 230 & $< 2.25$ & 3 \\
  IRAM 30m & 240 & 10.1 & 4 \\
  MIPS & 1875 & 377 & 5 \\
  IRAS & 3000 & 1370 & 6 \\
  MIPS & 4286 & 694 & 5 \\
  IRAS & 5000 & 390 & 6 \\
  \hline
 \end{tabular}
\label{tab:SED}
References -- (1)
 \citet{morganti06}; (2) \citet{nagar05}; (3) this paper; (4)
 \citet{wiklind95}; (5) \citet{temi07}; (6) \citet{moshir90}.
\end{table}

\section{Conclusions}
The interferometric observations of the CO(1-0) and CO(2-1) lines
presented here reveal a kiloparsec polar ring or disc of molecular gas
within the early-type galaxy NGC~2768. This molecular gas reveals the
potential of this galaxy to form kinematic substructure through star
formation. However, neither the kinematics nor the linestrengths of
NGC~2768 as measured by SAURON show clear evidence of a young stellar 
population. This limits any star formation coincident with the
molecular gas to be very recent and/or too small a luminosity-weighted
fraction of the background old stellar population to contribute to the
linestrengths and kinematics. The asymmetric and unrelaxed nature of
the CO(1-0) emission suggests that the
molecular gas in NGC~2768 was recently accreted and is possibly still
being accreted, in agreement with the much more extended but disturbed
H$\:${\small I} distribution. Searching the neighbourhood of NGC~2768
for potential gas-donor 
galaxies, many H$\:${\small I}-rich low-luminosity galaxies are found, all of
which could be the source of H$\:${\small I} seen in and around
NGC~2768, although UGC~4808 seems the most likely, being closest in 
projected distance and aligned with the extended H$\:${\small I} distribution. 

These CO observations of NGC~2768 probe new regions of parameter
space in terms of molecular gas in early-type galaxies, with recently
accreted, polar-rotating gas of only $6.8\times10^{7}$
M$_{\sun}$. As surveys have indicated lower CO detection rates in
higher-luminosity E and S0 galaxies \citep{lees91,sage07}, NGC~2768 is also notable for being
the brightest in B-band of the CO detections in \citet{CYB07}.
Whether the lack of observed star formation has to do with the (short)
time
since the gas has been accreted, the (low) mass of molecular gas available
for star formation, the polar dynamics of the gas, or some
host-galaxy property will take more investigation to discover, including
an analysis of the molecular gas distributions of other early-type galaxies with integral-field data.

\section*{Acknowledgements}
 We would like to thank Philippe Salome for help with
the reduction of the Plateau de Bure data and Marc Sarzi for useful
discussions. We are also grateful to
the SAURON Team for providing SAURON data as well as Tom Oosterloo for
providing the H$\:${\small I} datacube of NGC~2768. LMY acknowledges
support from grant NSF AST-0507432 and would like to thank the Oxford
Astrophysics Department for its hospitality during sabbatical work. We
would also like to thank the anonymous referee for helpful comments which
improved the paper.

The dust maps used observations made with the NASA/ESA Hubble
Space Telescope, obtained from the data archive at the Space Telescope
Institute. STScI is operated by the association of Universities for
Research in Astronomy, Inc. under the NASA contract  NAS 5-26555. 
The NASA/IPAC Extragalactic Database (NED) is operated by the Jet
Propulsion Laboratory, California Institute of Technology, under
contract with the National Aeronautics and Space Administration. This
research made use of HyperLEDA: http://leda.univ-lyon1.fr. The
Digitized Sky Surveys were produced at the Space Telescope Science
Institute under U.S. Government grant NAG W-2166. The images of these
surveys are based on photographic data obtained using the Oschin
Schmidt Telescope on Palomar Mountain and the UK Schmidt Telescope.

\label{lastpage}


\begin{thebibliography}{99}
\bibitem[\protect\citeauthoryear{Bournaud \&
    Combes}{2003}]{bournaud03}
  Bournaud F., Combes F., A\&A, 401, 817
\bibitem[\protect\citeauthoryear{Broeils \& van
    Woerden}{1994}]{broeils94}
  Broeils A. H., van Woerden H., 1994, A\&AS, 107, 129
\bibitem[\protect\citeauthoryear{Combes, Young \&
    Bureau}{Combes et al.}{2007}]{CYB07}
  Combes F., Young L.M., Bureau M., 2007, MNRAS, 377, 1795
\bibitem[\protect\citeauthoryear{de Vaucouleurs et
    al.}{1991}]{devaucouleurs91}
  de Vaucouleurs G., de Vaucouleurs A., Corwin H. G., Buta R. J.,
    Paturel G., Fouque P., 1991, Third Reference Catalog of Bright
    Galaxies, Vols. 1-3, XII. Springer-Verlag, Berlin (RC3)
\bibitem[\protect\citeauthoryear{de Zeeuw et al.}{2002}]{dezeeuw02}
  de Zeeuw T., et al., 2002, MNRAS, 329, 513
\bibitem[\protect\citeauthoryear{Devereux \&
    Hameed}{1997}]{devereux97}
  Devereux N. A., Hameed S., 1997, ApJ, 113, 599
\bibitem[\protect\citeauthoryear{Djorgovski \&
    Davis}{1987}]{djorgovski87}
  Djorgovski S., Davis M., 1987, ApJ, 313 59
\bibitem[\protect\citeauthoryear{Doi et al.}{2005}]{doi05}
  Doi A., Kameno S., Kohno K., Nakanishi K., Inoue M., 2005, MNRAS,
  363, 692
\bibitem[\protect\citeauthoryear{Dressler et al.}{1987}]{dressler87}
  Dressler A., Lynden-Bell D., Burstein D., Davies R. L., Faber S. M.,
  Terlevich R., Wegner G., 1987, ApJ, 313, 42
\bibitem[\protect\citeauthoryear{Emsellem et al.}{2001}]{emsellem01}
  Emsellem E., Greusard D., Combes F., Friedl D., Leon S., P\'econtal
  E., Wozniak H., 2001, A\&A, 368, 52
\bibitem[\protect\citeauthoryear{Emsellem et al.}{2004}]{emsellem04}
  Emsellem E., et al., 2004, MNRAS, 352, 721
\bibitem[\protect\citeauthoryear{Emsellem et al.}{2007}]{em07}
  Emsellem E., et al., 2007, MNRAS, 379, 401
\bibitem[\protect\citeauthoryear{Garcia}{1993}]{garcia93} Garcia
  A. M., 1993, A\&AS, 100, 47
\bibitem[\protect\citeauthoryear{Guilloteau \& Lucas}{2000}]{GL} 
  Guilloteau S., Lucas R., 2000, in Mangum J.G., Radford S.J.E., eds,
  ASP Conf. Ser. Vol. 217, Imaging at Radio through Submillimeter
  Wavelengths. Astron. Soc. Pac., San Francisco, p. 299
\bibitem[\protect\citeauthoryear{Hatch, Crawford \&
    Fabian}{2007}]{hatch}
  Hatch N. A., Crawford C. S., Fabian A. C., 2007, MNRAS, 380, 33
\bibitem[\protect\citeauthoryear{Haynes et al.}{1988}]{haynes88}
  Haynes M. P., Magri C., Giovanelli R., Starosta B. M., 1988, AJ, 95, 607
\bibitem[\protect\citeauthoryear{Heckman}{1980}]{heckman80}
  Heckman T. M., 1980, A\&A, 87, 152
\bibitem[\protect\citeauthoryear{Ho, Filippenko \& Sargent}{Ho et
    al.}{1997}]{ho97}
  Ho L. C., Filippenko A. V., Sargent W. L., 1997, ApJS, 98, 477
\bibitem[\protect\citeauthoryear{H\"ogbom}{1974}]{hogbom74}
  H\"ogbom J. A., 1974, A\&AS, 15, 417
\bibitem[\protect\citeauthoryear{Jesseit et al.}{2007}]{J}
  Jesseit R., Naab T., Peletier R. F., Burkert, A., 2007, MNRAS, 376, 997
\bibitem[\protect\citeauthoryear{Kawata, Cen \& Ho}{Kawata et
    al.}{2007}]{kawata07}
  Kawata D., Cen R., Ho L. C., 2007, ApJ, 669, 232
\bibitem[\protect\citeauthoryear{Kennicutt}{1998}]{kennicutt98}
  Kennicutt, R. C. Jr., 1998, ARA\&A, 36, 189
\bibitem[\protect\citeauthoryear{Kim}{1989}]{kim89}
  Kim D., ApJ, 346, 653
\bibitem[\protect\citeauthoryear{Knapp at al.}{1989}]{knapp}
  Knapp G.R., Guhathakurta P., Kim D., Jura M.A., 1989, ApJS, 70, 329
\bibitem[\protect\citeauthoryear{Komossa, B\"ohringer
    \& Huchra}{Komossa et al.}{1999}]{komossa99}
  Komossa S., B\"ohringer H., Huchra J.P., A\&A, 349, 88
\bibitem[\protect\citeauthoryear{Krips et al.}{2007}]{krips07}
  Krips M., et al., 2007, A\&A, 464, 553
\bibitem[\protect\citeauthoryear{Kuntschner et al.}{2006}]{kuetal06}
  Kuntschner H., et al., 2006, MNRAS, 369, 497 
\bibitem[\protect\citeauthoryear{Lees et al.}{1991}]{lees91}
  Lees J., Knapp G. R., Rupen M. P., Phillips T. G., 1991, ApJ, 379, 177
\bibitem[\protect\citeauthoryear{McDermid et al.}{2006}]{mcdermid06}
  McDermid R.M., et al., 2006, MNRAS, 373, 906
\bibitem[\protect\citeauthoryear{Morganti et al.}{2006}]{morganti06}
  Morganti R. et al., 2006, MNRAS, 371, 157
\bibitem[\protect\citeauthoryear{Moshir et al.}{1990}]{moshir90}
Moshir M. et al., 1990, IRAS Faint Source Catalog, version 2.0
\bibitem[\protect\citeauthoryear{Nagar, Falcke \& Wilson}{Nagar et
    al.}{2005}]{nagar05} Nagar N.M., Falcke H., Wilson A.S., 2005,
    A\&A, 435, 521
\bibitem[\protect\citeauthoryear{Nagar et al.}{2002}]{nagar02}
  Nagar N. M., Wilson A. S., Falcke H., Ulvestad J. S., Mundell C. G.,
  2002, ASPC, 258, 171
\bibitem[\protect\citeauthoryear{Okuda et al.}{2005}]{okuda05}
  Okuda T., Kohno K., Iguchi S., Nakanishi K., 2005, ApJ, 620, 673
\bibitem[\protect\citeauthoryear{Sage \& Welch}{2006}]{sage06}
  Sage L. J., Welch G. A., 2006, ApJ, 644, 850   
\bibitem[\protect\citeauthoryear{Sage, Welch \& Young}{2007}]{sage07}
  Sage L. J., Welch G. A., Young, L. M., 2007, ApJ, 657, 232
\bibitem[\protect\citeauthoryear{Sandage \& Bedke}{1994}]{sandage94}
  Sandage A., Bedke J., 1994, The Carnegie Atlas of Galaxies. DC:
  Carnegie Institute of Washington with The Flintridge Foundation,
  Washington DC (CAG)
\bibitem[\protect\citeauthoryear{Sarzi et al.}{2006}]{sarzi06}
  Sarzi M. et al. 2006, MNRAS, 366, 1151
\bibitem[\protect\citeauthoryear{Sault, Teuben \&
    Wright}{1995}]{sault95}
  Sault R. J., Teuben P. J., Wright M. C. H., 1995, ASPC, 77, 433
\bibitem[\protect\citeauthoryear{Schinnerer \&
    Scoville}{2002}]{schinnerer02}
  Schinnerer E., Scoville N., 2002, ApJ, 577, L103
  2004, AJ, 127, 2641 
\bibitem[\protect\citeauthoryear{Simien \& Prugniel}{1997}]{simien}
  Simien F., Prugniel P., 1997, A\&AS, 126, 519
\bibitem[\protect\citeauthoryear{Swaters}{1999}]{swaters99}
  Swaters R., 1999, PhD thesis, Rijksuniversiteit Groningen
\bibitem[\protect\citeauthoryear{Temi, Brighenti \& Matthews}{Temi et
    al.}{2007}]{temi07}
  Temi P., Brighenti F., Matthews W.G., 2007, ApJ, 660, 1215
\bibitem[\protect\citeauthoryear{Toomre}{1964}]{toomre64}
  Toomre A., 1964, SpJ, 139, 1217
\bibitem[\protect\citeauthoryear{Wozniak et al.}{2003}]{wozniak03}
  Wozniak H., Combes F., Emsellem E., Friedli D., 2003, A\&A, 409, 469
\bibitem[\protect\citeauthoryear{Wiklind \& Henkel}{1995}]{wiklind95}
  Wiklind T., Henkel C., 1995, A\&A, 297, L71
\bibitem[\protect\citeauthoryear{Young}{2002}]{young02}
  Young L. M., 2002, ApJ, 124, 788
\bibitem[\protect\citeauthoryear{Young}{2005}]{young05}
  Young L. M., 2005, ApJ, 634, 258 
\bibitem[\protect\citeauthoryear{Young, Bureau \& Cappellari}{Young et
  al.}{2008}]{young08}
  Young L. M., Bureau M., Cappellari M., 2008, ApJ, accepted
  (astro-ph/0712.4189)
	

  
\end{thebibliography}
\end{document}